\documentclass[aps, prx, noshowpacs, twocolumn, amsfonts, amsmath, amssymb, superscriptaddress,10pt]{revtex4-2}

\usepackage{xcolor,graphicx}
\usepackage[colorlinks=true]{hyperref}
\usepackage{soul}
\usepackage{multirow}

\newcommand{\equally}{These authors contributed equally.}
\newcommand{\poli}{Dipartimento di Fisica - Politecnico di Milano, p.za Leonardo da Vinci 32, 20133 Milano, Italy}
\newcommand{\ifn}{Istituto di Fotonica e Nanotecnologie - Consiglio Nazionale delle Ricerche (IFN-CNR), p.za Leonardo da Vinci 32, 20133 Milano, Italy}
\newcommand{\cnn}{Centre for Nanosciences and Nanotechnology, CNRS, Universit\'e Paris-Saclay, UMR 9001,10 Boulevard Thomas Gobert, 91120, Palaiseau, France}
\newcommand{\spz}{Dipartimento di Fisica, Sapienza Universit\`a di Roma, P.le Aldo Moro 5, 00185, Rome, Italy}
\newcommand{\quandela}{Quandela SAS, 7 Rue L\'eonard de Vinci, 91300 Massy, France}

\renewcommand{\vec}[1]{\underline{#1}}

\newtheorem{definition}{Definition}
\newtheorem{proposition}{Proposition}
\newtheorem{corollary}{Corollary}[proposition]

\newcommand{\ket}[1]{|#1\rangle}

\newcommand{\perm}{\mathrm{perm}\,}

\newcommand{\gtwo}{\mathit{g}^{(2)}}

\begin{document}


\title{Quantifying {\it n}-photon indistinguishability with a cyclic integrated  interferometer}
\thanks{Address correspondence to:\\Sarah E. Thomas: sarah.thomas@c2n.upsaclay.fr;\\ Andrea Crespi: andrea.crespi@polimi.it}

\author{Mathias Pont}
\thanks{\equally}
\affiliation{\cnn}

\author{Riccardo Albiero}
\thanks{\equally}
\affiliation{\poli}
\affiliation{\ifn}

\author{Sarah~E. Thomas}
\thanks{\equally}
\affiliation{\cnn}

\author{Nicol\`o Spagnolo}
\affiliation{\spz}

\author{Francesco Ceccarelli}
\affiliation{\ifn}

\author{Giacomo Corrielli}
\affiliation{\ifn}

\author{Alexandre Brieussel}
\author{Niccolo Somaschi}
\affiliation{\quandela}

\author{H\^elio Huet}
\author{Abdelmounaim Harouri}
\author{Aristide Lema\^itre}
\author{Isabelle Sagnes}
\author{Nadia Belabas}
\affiliation{\cnn}

\author{Fabio Sciarrino}
\affiliation{\spz}

\author{Roberto Osellame}
\affiliation{\ifn}

\author{Pascale Senellart}
\affiliation{\cnn}

\author{Andrea Crespi}
\affiliation{\poli}
\affiliation{\ifn}

\begin{abstract}
We report on a universal method to  measure the genuine indistinguishability of  $n$-photons -- a crucial parameter that determines the accuracy of optical quantum computing. Our approach relies on a low-depth cyclic multiport interferometer with $N = 2n$ modes, leading to a quantum interference fringe whose visibility is a direct measurement of the genuine $n$-photon indistinguishability. We experimentally demonstrate this technique for a 8-mode integrated interferometer fabricated using femtosecond laser micromachining 
and four photons from a quantum dot single-photon source. We measure a four-photon indistinguishability up to $0.81\pm 0.03$. 
This value decreases as we intentionally alter the photon pairwise indistinguishability. The low-depth and low-loss multiport interferometer design provides an efficient and scalable path to evaluate the genuine indistinguishability of resource states of increasing photon number. 
\end{abstract}


\maketitle


\section{Introduction}
\label{sec:intro}

Optical quantum computing has recently gained important momentum with the demonstration of  the so-called ``quantum advantage'' regime~\cite{brod2019p,zhong2020,zhong2021}. Quantum states of many indistinguishable photons are key elements for photonic quantum computing since the indistinguishable nature of the particles enables the implementation of effective photon-photon gates. Furthermore, scaling up to larger photon numbers is a key element to activate quantum speed up, implement fault tolerance, and reach a higher number of users in a quantum network. Experimentally, the efficient generation of identical single photons  is indeed an important technological challenge that is currently addressed with two main approaches, one based on multiplexing many heralded single-photon sources~\cite{Spring2017,Kaneda2019}, the other based on bright single emitters, namely  semiconductor quantum dots~\cite{senellart2017,Wang2019,Tomm2021, Thomas2021}.

On the fundamental side, quantifying the indistinguishability of the targeted multiple-photon states has recently been identified as a complex task. In the case of two photons, the indistinguishability is readily measured by performing a Hong-Ou-Mandel (HOM) interference experiment~\cite{HOM, menssen2017}. For two-photon wavepackets, the visibility of the quantum interference fringe gives access to the indistinguishability parameter, provided that the sources of noise are well understood~\cite{Fischer2018,Trivedi2020, Ollivier2021}.
However, when more than two photons are involved, the genuine indistinguishability of the full set of photons is not completely characterized by the HOM interference of all possible pairs \cite{menssen2017,giordani2020}, while only methods capable to provide lower and upper bounds to this quantity have been defined~\cite{brod2019w,giordani2020}.

Different generalizations of the HOM effect to the many-particle case have been proposed theoretically \cite{tichy2010,crespi2015,dittel2017,dittel2018,viggianiello2018} and observed in experiments \cite{crespi2016,viggianiello2018}. Such generalized HOM effects, which harness multi-port interferometers with specific symmetries, consist of suppression laws: indistinguishability of the photons in the input state results in a large number of forbidden output states among the many-particle possible ones.
While the observation of these suppression effects can be used to discriminate truly indistinguishable single-photon ensembles from other alternative input states \cite{tichy2014,viggianiello2018opt}, this has not been exploited to effectively quantify the genuine multi-photon indistinguishability. In fact, totally destructive interference may also be observed in case of partially-distinguishable photon states that still obey certain symmetries \cite{munzberg2021}. At present, only witnesses of indistinguishability have been proposed \cite{brod2019w,vandermeer2021}, while a general technique to quantitatively measure the genuine indistinguishability of many photons is missing.

Here, we propose a scalable way to quantify $n$-photon indistinguishability. Our scheme relies on a cyclic multi-port interferometer with $N=2n$ optical modes, composed of $2n$ beam splitters placed along two cascaded layers. Notably, this two-layer optical depth does not depend on $n$, thus keeping the layout extremely simple also for states with a large number of photons. We show that when $n$ indistinguishable photons are injected in the interferometer, the output distribution exhibits quantum interference depending on a single internal phase, and the interference visibility directly corresponds to the genuine $n$-photon indistinguishability. Conversely, no quantum interference is observed for input states with less than $n$ photons. We demonstrate experimentally our idea with 4 photons and an 8-mode version of this interferometer. The 4-photon state is obtained from a quantum dot single-photon source (QDSPS)~\cite{Somaschi2016} using a time-to-spatial mode demultiplexer, while the interferometer is fabricated in integrated optics by taking advantage of the three-dimensional capabilities of the femtosecond laser micromachining technology~\cite{meany2015,corrielli2021}. We observe full compatibility between the measured amplitude of the quantum interference fringe and the bounds given by the pair-wise indistinguishabilities of the 4-photon state. This is done for a large range of pairwise indistinguishability values of the input photons, which we tune through spectral filtering or exploiting the polarization degree of freedom. In addition, we see excellent agreement between our measurements and the numerical simulations that take into account the imperfections of our experimental apparatus.

\section{Theoretical proposal: cyclic interferometer design}

\subsection{Premise} 

As mentioned above, the typical way to measure the indistinguishability of two single photons exploits the HOM effect. Namely, the photons are injected in the two separate input ports of a balanced beam-splitter; the delay between the two photons is scanned by varying the optical length of one of the incoming paths, while coincidence detections at the two separate output ports are monitored. A dip in the interference pattern, i.e. the suppression of coincidence detection, is observed for null relative delay, and the visibility of this dip quantifies the indistinguishability of the photon pair.

As a matter of fact, this experimental layout is not the only one that gives access to the latter quantity. For instance, the two photons could be injected simultaneously in the separate input ports of a Mach-Zehnder interferometer with balanced arms, while again monitoring the coincidence detection of two photons at the two separate output ports. 
If the internal phase of the Mach-Zehnder interferometer is scanned, a quantum interference fringe is measured in the coincidences \cite{rarity1990}. This fringe shows half of the period that would be observed in the case of classical light, and its visibility is directly linked to the visibility of the HOM dip, thus also providing a quantification of the two-photon indistinguishability. 

Our proposed $2n$-mode cyclic interferometer (Fig.~\ref{fig:interf1}a) in some sense generalizes the latter kind of measurement to the case of $n$ photons. Indeed, we show that in our device $n$-photon interference fringes are measured while scanning one of the internal phase delays, and that the visibility of this fringe directly quantifies the $n$-photon indistinguishability. However, in contrast to the two-arm Mach-Zehnder interferometer, which demonstrates interference fringes with either single photons or classical light, in our case interference fringes cannot be observed with input states of less than $n$ photons. In fact, no closed Mach-Zehnder rings are present in our cyclic layout. 

In the following, we describe the interferometer layout and then discuss its properties in a series of Propositions. First, we show that the many phase delays that are involved in the optical circuit can be combined in a single relevant phase term (Section~\ref{sec:layout}). Then, we analyze theoretically multi-photon interference in this device, showing that a $n$-photon interference fringe is visible in certain conditions when varying the single phase term (Section~\ref{sec:interference}), and that the visibility of this fringe is a measure of the genuine $n$-photon indistinguishability (Section~\ref{sec:dist}).

\subsection{Layout of the proposed interferometer}
\label{sec:layout}

\begin{figure}
\centering
\includegraphics[scale=1]{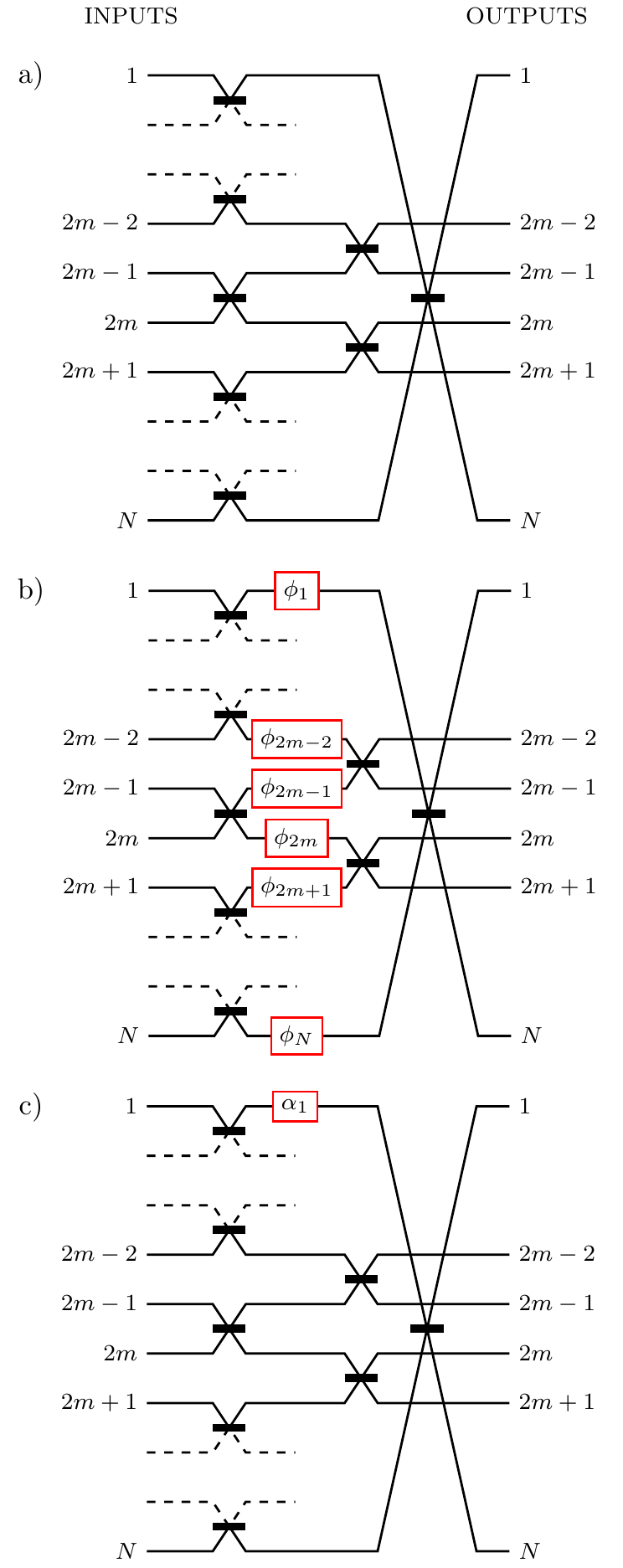}
\caption{\label{fig:interf1}(a) Schematic representation of  the proposed interferometer. (b) In most general terms, the  phase delays that are physically introduced by the optical propagation on the internal arms of the interferometers can be described by phase terms $\phi_i$. (c) Schematic representation of an interferometer equivalent to the one shown in panel b, where the many phase shifts $\phi_i$ are replaced by a single element $\alpha_1$ on the first arm, with phase delays considered as null in the other arms.}
\end{figure}

The proposed two-layer cyclic scheme is schematically represented in Fig.~\ref{fig:interf1}(a). The circuit contains $N=2n$ modes and is composed of two layers of cascaded beam-splitters. The beam-splitters in the first layer connect each odd mode with the subsequent even one (i.e., the mode $2m-1$ with the mode $2m$). The beam-splitters in the second layer connect each even mode with the subsequent odd one (i.e., the mode $2m$ with the mode $2m+1$). In this second layer, the $N$-th mode is connected with the first one, such that the device has perfect cyclical symmetry. All the beam splitters are balanced and symmetric.
The phase delays that correspond to the individual optical paths connecting the first to the second layer of beam splitters can be described by phase shifts with values $\phi_i$ (Fig.~\ref{fig:interf1}(b)). For the sake of simplicity, in the following we will refer to an interferometer with this layout just as Cyclic Interferometer (C.I.).

In general terms, the operation of a linear optical interferometer can be described by a unitary matrix $U$ that transforms, in the Heisenberg picture, the creation operators of the input modes into the creation operators of the output modes. This matrix allows to describe the interferometer operation on any possible input state.

When studying multi-photon interference, however, one is often interested in the operation of the linear circuit on Fock states. A generic Fock state of $k$ photons on $N$ modes can be written as:
\begin{equation}
\ket{s} = \left( \prod^k_{i=1} a^{\dagger}_{s_i} \right) \ket{0}
\label{eq:FockState}
\end{equation}
where $a^{\dagger}_{s_i}$ is the creation operator on the mode $s_i$. Such a state can be identified by the $k$-element vector $\vec{s} = \left(s_1, s_2, \ldots, s_k \right)$, with $1 \leq s_i \leq N$. 
In the following, we write as $P_{\vec g, \vec h}$ the probability to measure a certain Fock state $\vec h$ at the output when a given Fock state $\vec g$ is fed at the input of the device. 

\begin{definition} \label{def:equivalent}
Two $N$-mode interferometers are here defined as \emph{equivalent} if they give the same $P_{\vec g, \vec h}$ for each possible couple of input and output Fock states $\vec g$ and $\vec h$.
\end{definition}

Note that two equivalent interferometers may not be described by the same unitary matrix $U$. In particular, it is well known that two interferometers are equivalent in the sense of Definition~\ref{def:equivalent} if their optical circuits differ only in phase terms placed at the input or output ports~\cite{clements2016}. We refer the reader to the example in Appendix~\ref{sec:phases}, where the simple case of a Mach-Zehnder interferometer is analyzed.

\begin{proposition} \label{prop:equivalence}
The C.I. of Fig.~\ref{fig:interf1}b, with arbitrary phase delays $\phi_i$ in the internal arms, is \emph{equivalent} to the C.I. of Fig.~\ref{fig:interf1}c, where a single phase shift:
\begin{equation}
\alpha_1 = \sum_{m = 1}^{N/2} \phi_{2m-1} - \phi_{2m} \label{eq:alpha}
\end{equation}
is placed on the first arm, and null phase delays are assumed in the other ones.
\end{proposition}

The proof of this Proposition is given in full in Appendix~\ref{sec:simplification}. Such proof is constructed by showing that the phase delay $\phi_i$ on an arbitrary $i$-th arm can be replaced, without changing the unitary matrix $U$ of the interferometer, by: a null phase delay on the $i$-th arm, a different phase delay $\phi'_{i-1}$ on the $(i-1)$-th arm, and two additional phase shifts placed directly on the input or the output ports of the interferometer. This procedure can be iterated starting from the $N$-th arm up to the second one, and one ends up with a C.I. with several more phase shifts on the input and output ports, and null phase delays on all internal arms except the first one. In particular, one shows that the phase delay on the first arm now takes the value given by Eq.~\eqref{eq:alpha}. At this point, the added phases at the inputs and outputs can be removed producing a C.I. equivalent to the first one, which is indeed the one shown in Fig.~\ref{fig:interf1}c.

Actually, due to the cyclic symmetry of the device, one may alternatively implement the phase-delay replacements described above starting from a different point (say, the $(j-1)$-th arm), which results in a single remaining phase shifter placed on the $j$-th mode. Working out the details, the following Corollary is easily proven.
\begin{corollary} \label{coroll:equivalence}
There are $N$ C.I.s equivalent to the one shown in Fig.~\ref{fig:interf1}b, each with a phase element $\alpha_j$ on an arbitrary internal arm $j$ and null phases in the others. If $j$ is odd, $\alpha_j = \alpha_1$; otherwise $\alpha_j = - \alpha_1$.
\end{corollary}

For the experimenter, Proposition~\ref{prop:equivalence} and its Corollary~\ref{coroll:equivalence} mean that a single active phase shifter in the C.I. is sufficient to access all the operation space that would be given by controlling separately all the $\phi_i$.

\subsection{Interference of indistinguishable photons}
\label{sec:interference}

To study multiphoton interference in the C.I., we address first the case in which the device, having $N=2n$ modes, is injected with arbitrary input states having  $k \leq n-1$ photons. In fact, for these input states the following interesting fact holds.
\begin{proposition} \label{prop:lessPhotons}
If the C.I. is fed with $k \leq n-1$ photons, the values of the internal phases $\phi_i$ of the device do not have any influence on the output photon distribution.
\end{proposition}
Indeed, in this case, at least one of the beam splitters of the first layer has no photon at the input; say that this beam splitter is connecting modes $2m-1$ and $2m$. Because of Corollary~\ref{coroll:equivalence} we can compose an equivalent C.I. which has only one phase shift $\alpha_{2m}$ on the mode $2m$. No photon of the input state will then travel through this phase shift $\alpha_{2m}$, thus a change $\alpha_{2m}$ cannot affect the output distribution.  Since $\alpha_{2m}$ is defined as an algebraic sum of all the phases $\phi_i$, this means that variation of any $\phi_i$ does not affect the output.
{$\blacksquare$}

Note that, to prove Proposition~\ref{prop:lessPhotons}, we have crucially exploited the fact that interferometers that are equivalent according to Definition~\ref{def:equivalent} yield the same output distribution, and therefore we can investigate this distribution by choosing the most convenient device configuration among the possible ones.

Given the absence of closed interferometric rings in the C.I. layout, and the result of Proposition~\ref{prop:lessPhotons}, one may doubt that the internal phases $\phi_i$ influence the output distribution at all. Actually, interference fringes modulated by the $\phi_i$ can be observed in certain cases, in which the C.I. is fed with $n$ indistinguishable photons.

An instructive example is the case in which the $n$ identical photons are distributed one per odd input mode, and coincidence detection of one photon per odd output mode is monitored. We prove in Appendix~\ref{sec:permanents} the following:
\begin{proposition} \label{prop:prob135135}
Given a C.I. with $N=2n$ modes, and considering the input state $\vec g = (1,3,5,\dots,2n-1)$ of $n$ identical photons, the probability to detect the output state $\vec h = (1,3,5,\dots,2n-1)$ is given by:
\begin{equation}
P = \frac{1}{2^{2n-1}} \left( 1 + (-1)^n \cdot \cos \alpha_1 \right) \label{eq:Pind}
\end{equation}
\end{proposition}

This result indicates an interference fringe with unit visibility, as a function of the phase term $\alpha_1$. An intuition of the physical process underlying this interference fringe can be given by the observation that, due to the specific layout of the C.I., a photon from any given input mode $2m-1$ can reach only two different odd modes at the output: either the mode $2m-1$ (i.e. the same mode) or the mode $2m+1$ (i.e. the subsequent odd mode). It can be noted that, regardless of the distinguishable or indistinguishable nature of the input photons, there are only two possible evolutions that produce the desired output state $\vec h$: either every photon remains in the same odd mode in which it has been injected, or each photon shifts to the subsequent odd mode. In fact, if all the photons are perfectly indistinguishable, one cannot distinguish between the two possibilities and interference arises between the two evolution paths.

Actually, the case studied in Proposition~\ref{prop:prob135135} is the archetype of all the possible input and output states of $n$ photons that can result in an interference fringe as a function of the internal phases. Indeed we prove in Appendix~\ref{sec:states} the following result:
\begin{proposition} \label{prop:allStates}
Given a C.I. with $N=2n$ modes, an $n$-photon interference fringe as a function of the internal phases can be measured \emph{if and only if} the input state has precisely one photon per pair of input ports (1,2)-{\ldots}-($2m-1$,$2m$)-{\ldots}, and the output state has precisely one photon per pair of output ports (2,3)-{\ldots}-($2m$,$2m+1$)-{\ldots}-(1,$N$). In particular, for all input-output states of identical photons satisfying this rule, the detection probability takes the form:
\begin{equation}
P = \frac{1}{2^{2n-1}} \left(1 + (-1)^{n+p+q}\cdot \cos \alpha_1\right) \label{eq:PindFull}
\end{equation}
where $p$ is the number of occupied even modes in the input state, and $q$ is the number of occupied even modes in the output state.
\end{proposition}

For instance, in the case of an 8-mode interferometer, the combination $\lbrace \vec g = (2,3,5,7); \vec h = (1,2,4,7)\rbrace$ has one occupied even mode (the mode 2) in the input state $\vec g$ and two occupied even modes (the modes 2 and 4) in the output state $\vec h$. Since in this case $n=4$, one has $(-1)^{(4+1+2)}=-1$ and the term $\cos \alpha_1$ in the equation has to be taken with the minus sign.
 
We would like again to highlight the fact that, whereas Proposition~\ref{prop:prob135135} and \ref{prop:allStates} ensure that the internal phase $\alpha_1$ affects the quantum interference of $n$ photons, Proposition~\ref{prop:lessPhotons} warns that it does not have influence on two-photon experiments. This contrasts with the standard techniques currently in use to characterize multi-port devices, such as the one devised in Ref.~\cite{laing2012}, which allow to retrieve the unitary matrix of the interferometer  through one- and two-photon statistics only. These techniques are often tacitly assumed to be universal, but here they would not be able to provide the value of the phase $\alpha_1$. As a matter of fact, such methods implicitly assume that the unitary matrix of the device has at least one column and one row made of non-vanishing elements. This appears not to be the case here, where each input is connected with a small set of outputs, and thus the unitary matrix of the circuit is highly sparse.

\subsection{Interference of partially distinguishable photons}
\label{sec:dist}

We now consider the effect of input states of $n$ photons that are not perfectly indistinguishable. We describe these states as in Ref.~\cite{giordani2020}, using  a density matrix of the kind:
\begin{equation}
\rho = c_1 \rho^\parallel + \sum_i c_i \rho_i^\perp
\label{eq:decomp}
\end{equation}
where $\rho^\parallel$ is the state with all $n$ photons absolutely indistinguishable, while the states $\rho_i^\perp$, $i>1$, are states with at least two photons in mutually orthogonal states (i.e. distinguishable). The decomposition is convex, namely the real coefficients $c_i$ are all positive and $c_1 + \sum c_i = 1$.

In particular, we can first study again the $n$-photon experiment with input state $\vec g =  (1,3,5,\dots)$ and output state $\vec h =  (1,3,5,\dots)$. Note that, even if we use the same notation for $\vec g$ and $\vec h$ that we used for Fock states of identical photons, here the photons occupying the optical modes can be partly or fully distinguishable, and $\vec g$ is described by $\rho$.
\begin{proposition}
If the C.I. is fed with one photon per odd mode, with the $n$-photon state described by~\eqref{eq:decomp}, the probability to detect at the output again one photon per odd mode is given by:
\begin{equation}
P'=\frac{1}{2^{2n-1}}\left[ 1 + (-1)^n \cdot  c_1 \cos \alpha\right] \label{eq:pDistIndist}
\end{equation}
\end{proposition}

To calculate $P'$ we need to retrieve first the probability of the desired output state $\vec h$ considering as input $\rho^\parallel$ and  each of the components $\rho_i^\perp$.
The probability in the case of $\rho^\parallel$ is the one studied in Section~\ref{sec:interference}, and is given by Eq.~\eqref{eq:Pind}.
For any of the other states $\rho_i^\perp$, the fact that at least two photons of this state are orthogonal implies that at least two photons injected in adjacent odd modes are orthogonal, say $2m-1$ and $2m+1$. Considering the layout of the C.I, we observe that the distinguishability of these two photons implies the distinguishability of the two possible evolution paths that the photons follow in the interferometer after the first layer of beamsplitter: since we can distinguish the photon initially on mode $2m-1$ from the photon initially on mode $2m+1$, we can tell whether they have remained on the same output port or they have shifted two modes forwards (see the discussion following  Proposition~\ref{prop:prob135135}, in the previous Section). Indeed, if we know the path followed by one photon (e.g. the $(2m-1)$-th one), then there is only one possible path for all the others, independently of their distinguishability or indistinguishability, that produces an output state with photons on odd modes. 
This leads to conclude that, if at least one photon is distinguishable from the others, then we can track the path of all of them and they behave as if they were all distinguishable. 

The probability of detecting an output state $\vec h =  (1,3,5,\dots)$ for input photons that are all distinguishable and in a state $\vec g =  (1,3,5,\dots)$, is demonstrated explicitly in Appendix~\ref{sec:permanents} to be: 
\begin{equation}
P_\mathrm{dist} =\frac{1}{2^{2n-1}}
\label{eq:pOddDist}
\end{equation}
Because of the above discussion, this expression holds for all the states $\rho_i^\perp$.

Finally, the probability of detecting one photon per each odd output mode, for an input state with one photon per each odd mode, described by  Eq.~\eqref{eq:decomp}, is computed as:
\begin{align}
P' &= c_1 P + \sum_i c_i P_{dist} = c_1 P + (1-c_1) P_{dist} =\notag\\
&=\frac{1}{2^{2n-1}}\left[ 1 + (-1)^n \cdot c_1 \cos \alpha\right]
\end{align}
which proves the Proposition. $\blacksquare$

One notes that Eq.~\eqref{eq:pDistIndist} describes a multi-photon interference fringe with visibility $\mathcal{V} = c_1$.
Therefore such fringe visibility is a direct indicator of the genuine $n$-indistinguishability of the set of $n$ input photons.

It is not difficult to show that the above discussion also applies identically to all the other pairs of $n$-photon input and output states which give an observable interference fringe for varying $\alpha_1$, as described in Proposition~\ref{prop:allStates}. 
Hence, if $n$ photons are distributed  with one photon for each pair of ports (1,2)-{\ldots}-($2m-1$,$2m$)-{\ldots}, one can characterize their $n$-photon indistinguishability by monitoring the interference fringe observable for any of the output states having one photon for each pair of ports (2,3)-{\ldots}-($2m$,$2m+1$)-{\ldots}-(1,$N$).

\section{Experimental results}
\label{sec:experimental}

\subsection{Measurement of 4-photon indistinguishability}
\label{sec:4exp}

\begin{figure*}
    \centering
    \includegraphics[width=\linewidth]{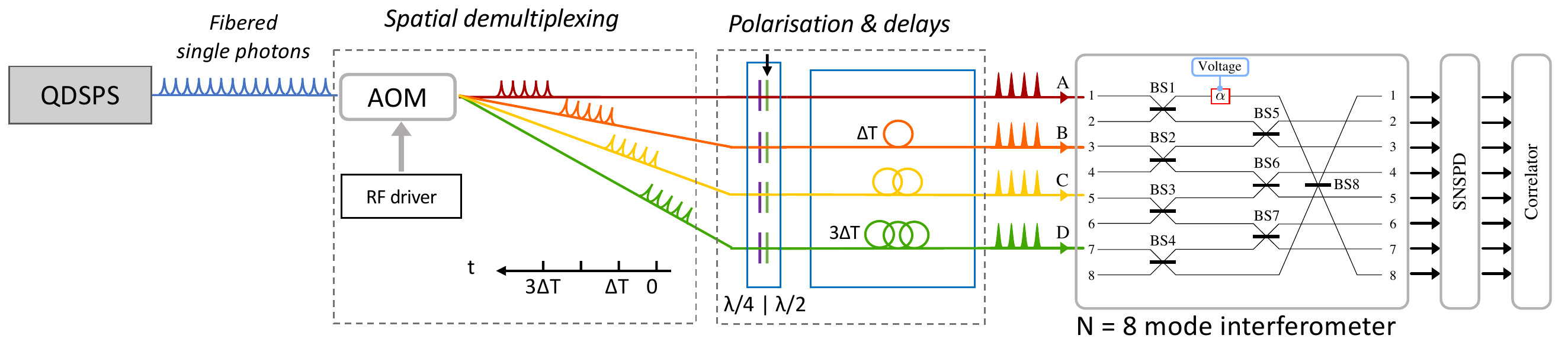}
    \caption{\label{fig:exp_setup} Schematic of the experimental setup. A quantum dot based single-photon source (QDSPS) is excited periodically and emits a stream of single photons. The photons are separated into four spatial modes using a demultiplexer which is based on a periodically driven acousto-optic modulator (AOM). The polarization of the photons is controlled using waveplates, and delay fibers are used to ensure the simultaneous arrival times of the four photons. The photons are sent into the odd modes of the integrated C.I., fabricated by femtosecond laser micromachining. The photons are detected at the output of the interferometer using superconducting nanowire single-photon detectors (SNSPDs), and four-photon coincidences are recorded using a correlator. The internal phase of the interferometer $\alpha_1$ is varied by scanning the voltage applied to the microheater realized on the chip surface.} 
\end{figure*}

We demonstrate experimentally the use of an 8-port C.I. to characterize the indistinguishability of 4-photon states. The C.I. is realized as a reconfigurable waveguide circuit, using the femtosecond laser direct writing technology~\cite{meany2015,flamini2015,ceccarelli2019,corrielli2021}. In detail, single-mode waveguides are inscribed in a commercial alumino-borosilicate glass substrate (Eagle XG, Corning Inc.) by irradiating the desired paths with multiple laser scans and by performing thermal annealing thereafter \cite{arriola2013}.Note that, to implement the C.I. layout in a compact fashion, waveguides are required to pass one over the others without crossing, to reach the last beam-splitter connecting modes 1 and $N$; this is uniquely made possible by the three-dimensional capabilities of femtosecond laser micromachining. The integrated circuit has an overall size of about 1$\times$30~mm$^2$.
A phase shifter is fabricated on the top surface, in corrispondence of one of the interferometer's arm, by patterning a resistive microheater in a gold film through femtosecond laser ablation. Driving an electric current in the microheater produces a local temperature increase, which in turn induces a controlled phase shift on the underlying waveguides due to the thermo-optic effect. To interface the circuit with the photon sources and with the detectors, single-mode fibers are glued with UV-curing resin, to both the input and the output ports of the chip. 
In order to enable the efficient collection of four-photon coincidences, particular care was given to reducing the insertion losses of the optical device, which result lower than 2 dB.

A four-photon state is generated from a quantum dot single-photon source (QDSPS). The QDSPS is based on a neutral InGaAs quantum dot embedded in an electrically-contacted micropillar cavity~\cite{Nowak2014a,Somaschi2016}, which is placed inside a cryostat at 5~K. The QDSPS is excited using acoustic-phonon-assisted near-resonant excitation~\cite{Thomas2021} using a 15~ps laser pulse centred at 924.4~nm, at a repetition rate of 82~MHz. This excitation scheme allows for increased source efficiency and stability, a key requirement for multi-photon experiments. 
The single photons are emitted at 925.0~nm and are separated from the excitation laser using three 0.8~nm bandpass filters with a transmission of 95\% and a laser suppression of approximately 40~dB per filter, before being collected into a single-mode fiber. We characterise the single-photon purity of the source in a Hanbury-Brown and Twiss measurement, obtaining $\mathcal{P} = 1-g^{2}(0) = (98.1\pm0.1)\%$. 

The stream of single photons is separated into four spatial modes using a demultiplexer (DMX) from Quandela. The DMX consists of an acousto-optic modulator (AOM) which, when driven by a time-varying radio-frequency (RF) signal, diffracts the incoming light into different spatial modes depending on the driving frequency. The time-varying signal driving the AOM has a total cycle time of $4 \times \Delta T$: the RF signal is ON at three distinct frequencies for $\Delta T$ each, and then OFF for $\Delta T$.
As pictured in Fig.~\ref{fig:exp_setup} the incoming photons are thus switched among four spatial modes: three corresponding to first-order diffraction through the AOM at different frequencies, and one corresponding to direct transmission through the AOM when no RF signal is applied. Note that the frequency shift of the photons induced via first-order diffraction through the AOM  is significantly smaller than the bandwidth of the photons ($\sim$200~MHz versus $\sim$2.1~GHz), and hence it does not affect their indistinguishability. After the DMX, the photons in all four spatial modes are coupled into single-mode fibers of different lengths such that the photons enter the C.I. simultaneously. We can experimentally achieve a precision of a few picoseconds in the arrival time of the single photons, which is much smaller than their lifetime of about 145~ps.
A quarter and a half waveplate are used on each spatial mode to adjust the polarization of each photon at the input of the C.I. Special care was taken to achieve passive mechanical and thermal stability of the DMX and fiber delays. We use two versions of the DMX with different cycle times $4\,\Delta T$ (see Section \ref{sec:exp_dist}), and the total transmission of the DMX is $\sim45\%$ ($\sim65\%$) for the first (second) design.

The initial design of the DMX has an operating time of $\Delta T = 320$~ns, meaning that the photons entering ports 1 and 7 of the interferometer were emitted by the QDSPS at a separation time of 960~ns. In order to understand the impact {of delays} on the multi-photon state, we investigate the indistinguishability of the photons emitted by the QDSPS at different separation times. We measure the indistinguishability via two-photon HOM interference in a path-unbalanced Mach-Zender interferometer. We extract the visibility of HOM interference, $V_\mathrm{HOM}$, and correct for the non-zero $g^{(2)}(0)$ to extract the indistinguishability of the single photon component, $M_\mathrm{s}$~\cite{Ollivier2021}. When the delay between the photons is $12.3$~ns, corresponding to the pulse separation of the laser, the visibility (indistinguishability) is $V_\mathrm{HOM} = (88.6\pm0.1)\%$ ($M_\mathrm{s} = (92.3\pm0.1)\%$). At longer separation times the visibility (indistinguishability) decreases due to residual charge noise in the QD environment which leads to spectral wandering of the QD emission, as shown in Table~\ref{tab:VvsT} (Appendix~\ref{app:exp}), and at a delay of 960~ns the HOM visibility (indistinguishability) is $V_\mathrm{HOM} = (72.7\pm0.1)\%$  ($M_\mathrm{s} = (76.0\pm0.1)\%$). We note that such decrease -- stronger than previously reported~\cite{Loredo2016} -- arises from the specific doping structure of the devices under investigation, which favours a particular charge state of the quantum dot at the expense of increased charge noise.  

\begin{table}[tp]
\begin{ruledtabular}
\begin{tabular}{cc|cc}
\textbf{Output State} & \textbf{Fringe sign} & \textbf{Output State} & \textbf{Fringe sign}\\
\hline
(1, 3, 5, 7) & + & (1, 3, 5, 6) & - \\
(1, 3, 4, 6) & + & (1, 3, 4, 7) & - \\
(1, 2, 5, 6) & + & (1, 2, 5, 7) & - \\
(1, 2, 4, 7) & + & (1, 2, 4, 6) & - \\
(3, 5, 6, 8) & + & (3, 5, 7, 8) & - \\
(3, 4, 7, 8) & + & (3, 4, 6, 8) & - \\
(2, 5, 7, 8) & + & (2, 5, 6, 8) & - \\
(2, 4, 6, 8) & + & (2, 4, 7, 8) & - \\
\end{tabular}
\end{ruledtabular}
\caption{\label{tab:states}Output states whose detection probability is described by Eq.~\eqref{eq:PindFull} when four identical photons in the state (1, 3, 5, 7) are injected at the input of the C.I. The sign of the interference fringe, i.e. the sign of the term $\cos \alpha_1$, is specified for each state.}
\end{table}

\begin{figure*}
    \centering
    \includegraphics[width=0.9\linewidth]{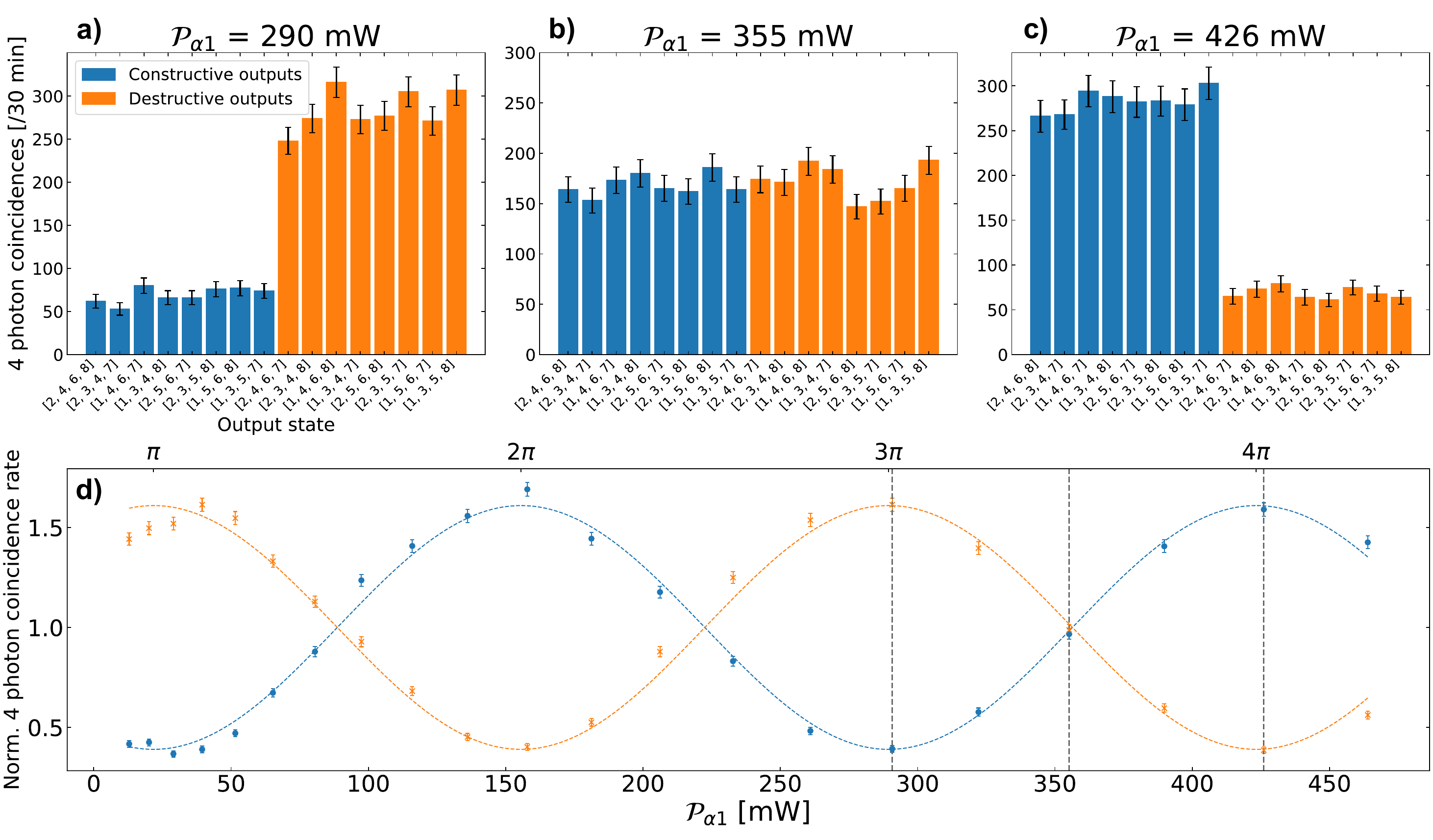}
    \caption{\label{fig:histo_indist}(a-c) Number of detected coincidences in 30 minutes for the 16 four-photon output states for a given electrical power $P_{\alpha}$ applied to the internal phase shifter of the integrated interferometer. We observe a clear shift from (a) destructive interference at $P_{\alpha}=290$~mW ($\alpha_1 \sim 3\pi$) to (c) constructive interference at $P_{\alpha}=426$~mW ($\alpha_1 \sim 4\pi$). The error bars indicate the shot-noise error $\sqrt{N}$ for a given number of coincidences $N$. (d) The total normalized 4-photon coincidence rate (sum of all eight output states) for the constructive and destructive outputs, as a function of $P_{\alpha1}$ or the phase $\alpha_1$.} 
\end{figure*}

We inject four simultaneous photons with parallel polarization into the input ports (1,3,5,7) of the interferometer, and  monitor the four-photon coincidences on the output. Table~\ref{tab:states} gives the 16 possible four-photon output states where we expect to observe an interference fringe according to the discussion in Section~\ref{sec:interference}. The fringe, i.e. the term $\cos \alpha_1$ in the detection probability, takes either a plus or minus sign according to Proposition~\ref{prop:allStates}, as indicated in  Table~\ref{tab:states}.
Figure~\ref{fig:histo_indist}(a)-(c) show the number of detected four-photon coincidence events in 30 minutes in each of the 16 output states, and we see a clear difference in the detected probability for the states that show constructive and destructive interference. Each histogram is for a different value of the electric power $\mathcal{P}_{\alpha_1}$ dissipated on the resistive microheater, which changes the internal phase $\alpha_1$. The total four-photon coincidence rate here is approximately 1.6~Hz. The integration window to define a 4-photon coincidence is set to 1~ns. With this integration window the accidental-to-coincidences ratio is in the order of $0.013\pm0.002$; this is defined as the ratio between the average number of 4-photon coincidences detected at the output with 3 photons sent into the interferometer, and the average number of 4-photon coincidences detected when 4 photons are injected.
We add together the detected events of all 8 states characterized by fringes with the same sign, to increase the measurement statistics. In Fig.~\ref{fig:histo_indist}(d) we plot the total number of coincidences for the outputs corresponding to constructive (blue) or destructive (orange) interference as a function of $\mathcal{P}_{\alpha_1}$, normalized to the mean value.
We observe a clear interference fringe in the four-photon coincidences, whilst no variation is observed in the two- and three-photon coincidence rates (graphs reported in Appendix \ref{app:2&3-coincidences}). 
We fit the data according to $P' \propto (1 \pm c_1 \cos \alpha)$ for both the constructive and destructive interference fringes. We extract a visibility, or equivalently a four-photon indistinguishability, of $c_1 = 0.61 \pm 0.01$.

\begin{figure*}
    \centering
    \includegraphics[width=0.9\linewidth]{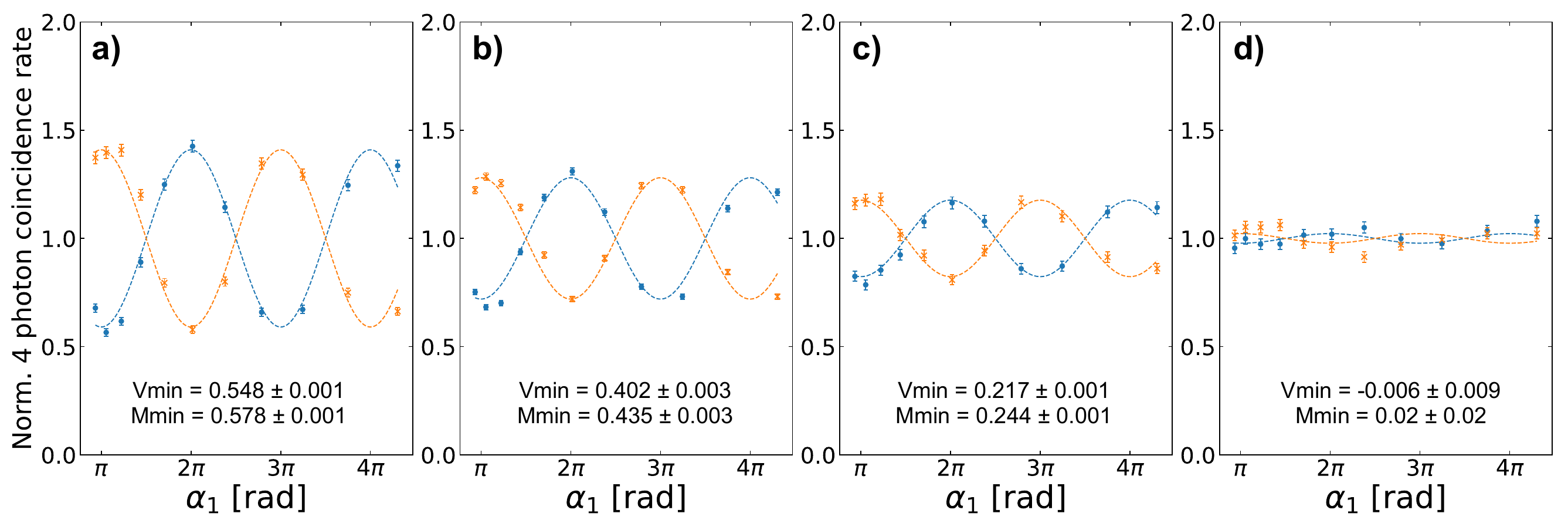}
    \caption{The polarization of photon A is rotated using a half waveplate to make it more distinguishable from the other photons. (a)-(d) The measured 4-photon interference fringe when photon A is made increasingly distinguishable. The visibility of the interference fringe decreases as the minimum measured two-photon Hong-Ou-Mandel interference visibility (extracted single-photon indistinguishability), $V_\mathrm{min}$ ($M_\mathrm{min}$) decreases. The measured 4-photon interference fringe contrast drops to almost zero when photon A is almost fully distinguishable.} 
    \label{fig:c1_Vmin}
\end{figure*}

\subsection{Verification: two-photon bounds and full model}
\label{sec:exp_model}

In order to verify and qualify this experimental result, we use two techniques. Firstly, we deduce an upper and a lower bound on the four-photon indistinguishability using the two-photon pairwise indistinguishabilities. The geometry of this interferometer also allows us to simultaneously measure some of the pairwise indistinguishabilities of the input photons, without changing the experimental set-up. In particular, photons A and B undergo HOM interference at beam splitter BS5 (see Fig.~\ref{fig:exp_setup}). Therefore, by observing the two-photon coincidences on outputs 2 and 3 we  directly measure the HOM interference visibility (and the two-photon indistinguishability~\cite{Ollivier2021}) between photons A and B, $V_\mathrm{AB}$ ($M_\mathrm{AB}$). Similarly, the interference between photons B and C is measured at outputs 4 and 5, C and D at outputs 6 and 7, and D and A at outputs 1 and 8. Note that these measurements can be carried out simultaneously while monitoring also the four-photon coincidences.

We then use these values to obtain upper and lower bounds for $c_1$, by following the methodology discussed in Ref.~\cite{giordani2020}. In particular, by considering the specific layout of the C.I., the following bounds for $c_1$, based on the four indistinguishabilities $(M_\mathrm{AB},M_\mathrm{BC},M_\mathrm{CD},M_\mathrm{DA})$ that are experimentally accessible, hold (see Appendix~\ref{app:bounds}):
\begin{align}
    c_1 & \geq M_\mathrm{AB}+M_\mathrm{BC}+M_\mathrm{CD}+M_\mathrm{DA}-3 \label{eq:c1_lower_bound}\\
    c_1 & \leq \mathrm{min} \left( M_\mathrm{AB},M_\mathrm{BC},M_\mathrm{CD},M_\mathrm{DA} \right) \label{eq:c1_upper_bound}
\end{align}
For the data presented in Fig.~\ref{fig:histo_indist}, the measured pairwise HOM visibilities (indistinguishabilities) are $V_\mathrm{AB} = 0.727 \pm 0.001$ ($M_\mathrm{AB} = 0.760 \pm 0.002$), $V_\mathrm{BC} = 0.790 \pm 0.001$, ($M_\mathrm{BC} = 0.825 \pm 0.002$), $V_\mathrm{CD} = 0.848 \pm 0.001$, ($M_\mathrm{CD} = 0.884 \pm 0.002$), $V_\mathrm{DA} = 0.755 \pm 0.002 $, ($M_\mathrm{DA} = 0.789 \pm 0.003 $). 

In order to establish robust upper and lower bounds on $c_1$ we account for the experimental error in the measured indistinguishability values via a bootstrapping approach (see Appendix~\ref{app:bounds}). This provides upper and lower bounds on the four-photon indistinguishability of $0.235 \leq c_1 \leq 0.766$. The measured value  $c_1 = 0.61 \pm 0.01 $ falls indeed within these bounds.

Secondly, we developed a model to simulate the experiment and explore the effect of experimental imperfections. In fact, the visibility of the multi-photon interference precisely corresponds to the parameter $c_1$, which is the multi-photon indistinguishability coefficient in the photonic density matrix of Eq.~\eqref{eq:decomp}, only in the ideal situation outlined in Section~\ref{sec:dist}. However, in a real experiment other physical effects can be responsible for a reduction in the measured visibility value. Some of these imperfections are intrinsically due to the source properties, such as multi-photon emission, while other ones may depend on the measurement apparatus, and include losses, imperfections of the fabricated interferometer, and unbalanced detection efficiencies. We developed a complete model to evaluate how such imperfections affect the measurement of the visibility and the extracted value of $c_1$ (see Appendix~\ref{app:model} for more details). By using the actual parameters of the apparatus, the model predicts a value of $c_1 = 0.590 \pm 0.005$, which is in good agreement with the experimental results. This value increases to $c_1 = 0.661 \pm 0.006$ when only the partial photon distinguishability is included (deduced from the four accessible two-photon HOM measurements), while all other noise sources are neglected. This means that all other sources of imperfections contribute to a reduction of $\sim 0.07$ in the measured value of $c_1$. More specifically, the main contribution to this reduction in the present experiment is found to be provided by multi-photon emission ($g^{(2)}(0) \simeq 0.019 \neq 0$), while the effects of circuit errors and unbalanced detection efficiencies are negligible. This additional analysis allows us to evaluate the impact of each experimental imperfection, and provides a detailed benchmark for the performance of the experiment and photon source. 

\begin{figure}
    \centering
    \includegraphics[width=\linewidth]{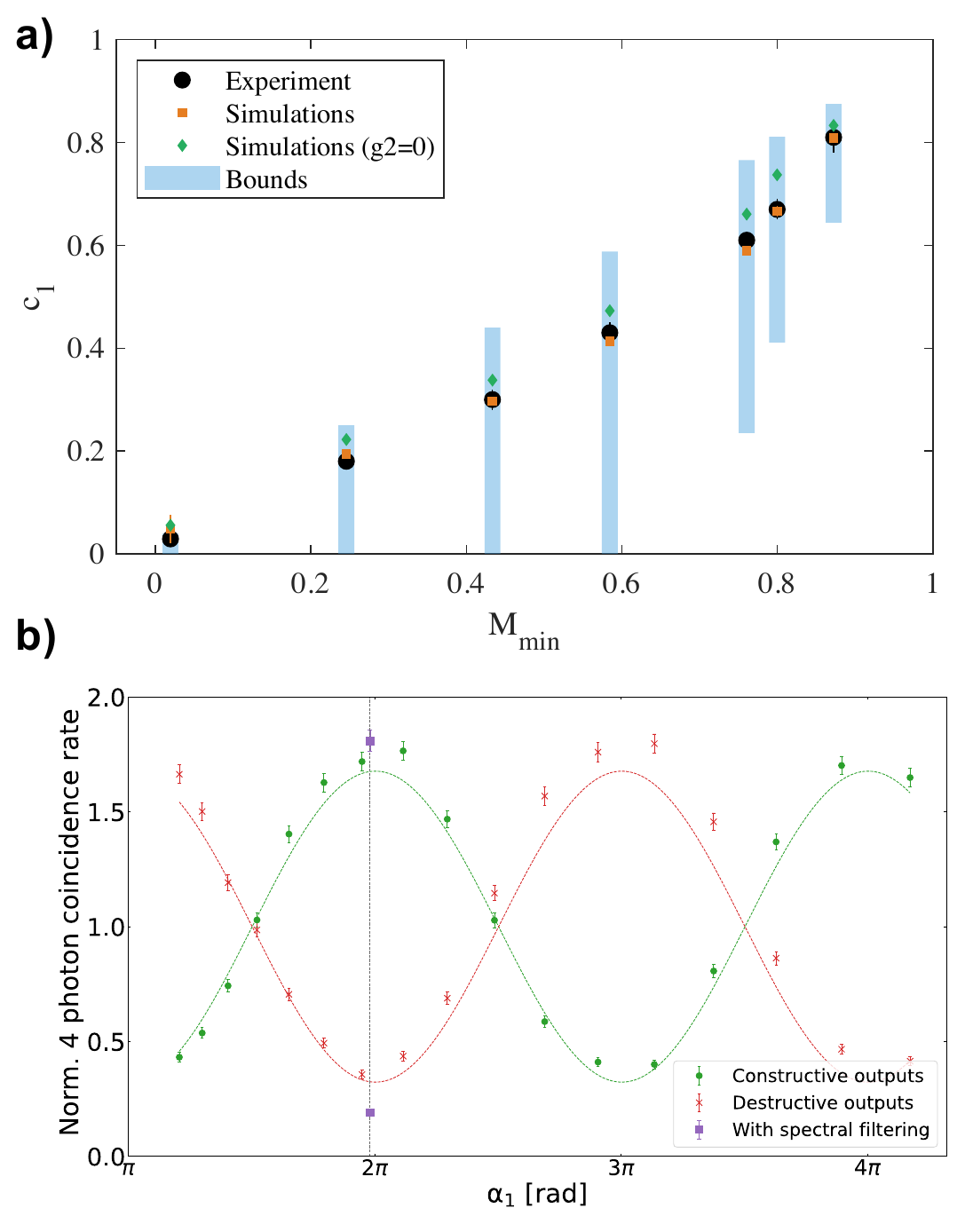}
    \caption{(a) Four-photon indistinguishability, $c_1$, as a function of the minimum pairwise indistinguishability, $M_\mathrm{min}$. Black circles: experimental data, extracted from the visibility of the four-photon interference fringe. Orange squares: simulated interference visibility including all experimental imperfections. Green diamonds: simulated interference visibility only accounting for partial distinguishability of the input photons, and assuming zero multi-photon emission ($g^{(2)} = 0$), an ideal interferometer, and perfect detection efficiency. Blue shaded regions: upper and lower bounds calculated from Eqs.~\eqref{eq:c1_lower_bound}-\eqref{eq:c1_upper_bound} using the four measured pairwise indistinguishabilities. (b) Measured four-photon coincidence fringe as a function of the internal phase of the interferometer, $\alpha$, using a faster DMX to {give access to} a higher $M_\mathrm{min} = 0.800 \pm 0.004$. The visibility here is $c_1  = 0.67 \pm 0.02$. The purple points were obtained using a {12 pm }Fabry-Perot etalon to spectrally filter the photons and improve the minimum pairwise indistinguishability to  $M_\mathrm{min} = 0.8727 \pm 0.0007$. Only one phase value is sampled here, and we obtain a visibility of $c_1 = 0.81 \pm 0.03$. }
    \label{fig:Simulations_Bounds}
\end{figure}
\subsection{Tuning Indistinguishability}
\label{sec:exp_dist}

We measure how the visibility of the interference fringe varies when one of the photons is made gradually distinguishable. We plot in Fig.~\ref{fig:c1_Vmin}a-d the four-photon interference fringes as we make photon A more and more distinguishable by rotating its polarization with a half waveplate (see arrow in Fig.~\ref{fig:exp_setup}).
For each data set we fit the interference fringe according to $P' \propto 1 \pm c_1 \cos \alpha_1$ and extract $c_1$. The experimental values of $c_1$ are plotted as a function of the minimum measured pairwise indistinguishability $M_\mathrm{min}$ in Fig.~\ref{fig:Simulations_Bounds}a as black circles. We calculate the upper and lower bounds for the expected value of $c_1$ from the two-photon overlaps, and these are shown as the shaded blue regions in the same graph. We also perform numerical simulations which account for experimental imperfections, as described before and more fully in Appendix~\ref{app:model} and calculate the expected visibilities, which are plotted as orange squares.
In all cases, there is a very good agreement between the data and simulations. For comparison, we calculate the expected value of the visibility if only the partial indistinguishability of the photons was considered (i.e. a perfect interferometer and $\gtwo = 0$), and these values are plotted as green diamonds.

We finally increase the two-photon indistinguishability using two different methods, in order to explore higher values of $M_\mathrm{min}$ and $c_1$. Firstly we use the second version of the DMX which has a shorter operating time of $\Delta T = 175$~ns. The maximum separation time between the interfering photons is then 525~ns and therefore the minimal two-photon indistinguishability is higher at $M_\mathrm{min} = 0.800 \pm 0.004$, because of reduced spectral wandering of the quantum dot between photon emission. In this case we extract a four-photon indistinguishability of $c_1 = 0.67 \pm 0.02$.
The experimental interference fringes are shown in Fig.~\ref{fig:Simulations_Bounds}b, and the extracted value of $c_1$ is also plotted in Fig.~\ref{fig:Simulations_Bounds}a together with the calculated bounds and simulated values. Secondly, we add a 12~pm Fabry-Perot etalon to spectrally filter the single-photon source. This decreases  the effect that spectral wandering has on the two-photon indistinguishability by post-selecting the single-photon emission within a narrow bandwidth. However, this reduces the source brightness by approximately at factor of 3 which significantly decreases the four-photon coincidence rate, thus we characterize this condition only for $\alpha_1 = 2\pi$ (see the purple dots in Fig.~\ref{fig:Simulations_Bounds}b, which were acquired with an integration time of 16 hours).
The minimal two-photon indistinguishability in the latter case is $M_\mathrm{min} = 0.8727 \pm 0.0007$ and the experimental four-photon indistinguishability is $c_1 = 0.81 \pm 0.03$. As can be seen in Fig.~\ref{fig:Simulations_Bounds}(a) there is again a good agreement between the experiment and simulations for these data points. 

\section{Conclusions}
\label{sec:conclusion}

We have proposed a scalable and robust way to measure the genuine $n$-photon indistinguishability of multi-photon states. Our method relies on a  photonic circuit with 2$n$ modes and a cyclic array of beam splitters. When $n$ photons are injected in the circuit in specific input configurations, a quantum interference fringe is observed while scanning the internal phases of the device. In practice, we demonstrate that it is sufficient to harness a single phase term, governed by a single thermo-optic phase shifter fabricated upon one arm of the interferometer. The visibility of this fringe provides a direct quantification of the multi-photon indistinguishability.

Our experimental study with 4-photons demonstrates the reliability of our approach. In fact, the conducted measurements are shown to be very robust against the interferometer imperfections and allow access to the true multi-photon indistinguishability for photons of various quantum purity. 

We emphasize that our C.I. can be scaled to an arbitrary number of modes, while keeping the same depth of only two layers of beam splitters and thus maintaining low optical loss. This makes it a powerful tool  to characterise multi-photon indistinguishability for increasing number of photons, a key resource in photonic quantum technologies. 

We believe this study may open new paths in fundamental research on quantum interference beyond photonics. Indeed, the novel design of our device shows that, by interfering cyclically a set of quantum particles, it is possible to access physical properties of the whole such as their genuine quantum indistinguishability. Further ramifications may be foreseen in the quantum metrology field. In fact, we provide here an example of device whose properties (namely, its internal phases) are accessible and measurable only with states composed by a minimum number of photons, while they are transparent to Fock states with lower numbers.

\begin{acknowledgments}
This work is partly  supported by the European Union's Horizon 2020 research and innovation programme under the PHOQUSING project GA no. 899544, by the European Union's Horizon 2020 Research and Innovation Programme QUDOT-TECH under the Marie Sklodowska-Curie Grant Agreement No. 861097, by the IAD-ANR support ASTRID program Projet ANR-18-ASTR-0024 LIGHT, by the QuantERA ERA-NET Cofund in Quantum Technologies project HIPHOP, by the French RENATECH network, by the Paris Ile-de-France R\'{e}gion in the framework of DIM SIRTEQ. 

The interferometer fabrication was partially performed at PoliFAB, the micro- and nanofabrication facility of Politecnico di Milano. The authors would like to thank Mr. Cl\'ement Gouriou for stimulating discussions.
\end{acknowledgments}


\begin{thebibliography}{44}%
\makeatletter
\providecommand \@ifxundefined [1]{%
 \@ifx{#1\undefined}
}%
\providecommand \@ifnum [1]{%
 \ifnum #1\expandafter \@firstoftwo
 \else \expandafter \@secondoftwo
 \fi
}%
\providecommand \@ifx [1]{%
 \ifx #1\expandafter \@firstoftwo
 \else \expandafter \@secondoftwo
 \fi
}%
\providecommand \natexlab [1]{#1}%
\providecommand \enquote  [1]{``#1''}%
\providecommand \bibnamefont  [1]{#1}%
\providecommand \bibfnamefont [1]{#1}%
\providecommand \citenamefont [1]{#1}%
\providecommand \href@noop [0]{\@secondoftwo}%
\providecommand \href [0]{\begingroup \@sanitize@url \@href}%
\providecommand \@href[1]{\@@startlink{#1}\@@href}%
\providecommand \@@href[1]{\endgroup#1\@@endlink}%
\providecommand \@sanitize@url [0]{\catcode `\\12\catcode `\$12\catcode
  `\&12\catcode `\#12\catcode `\^12\catcode `\_12\catcode `\%12\relax}%
\providecommand \@@startlink[1]{}%
\providecommand \@@endlink[0]{}%
\providecommand \url  [0]{\begingroup\@sanitize@url \@url }%
\providecommand \@url [1]{\endgroup\@href {#1}{\urlprefix }}%
\providecommand \urlprefix  [0]{URL }%
\providecommand \Eprint [0]{\href }%
\providecommand \doibase [0]{https://doi.org/}%
\providecommand \selectlanguage [0]{\@gobble}%
\providecommand \bibinfo  [0]{\@secondoftwo}%
\providecommand \bibfield  [0]{\@secondoftwo}%
\providecommand \translation [1]{[#1]}%
\providecommand \BibitemOpen [0]{}%
\providecommand \bibitemStop [0]{}%
\providecommand \bibitemNoStop [0]{.\EOS\space}%
\providecommand \EOS [0]{\spacefactor3000\relax}%
\providecommand \BibitemShut  [1]{\csname bibitem#1\endcsname}%
\let\auto@bib@innerbib\@empty
\bibitem [{\citenamefont {Brod}\ \emph
  {et~al.}(2019{\natexlab{a}})\citenamefont {Brod}, \citenamefont {Galv{\~a}o},
  \citenamefont {Crespi}, \citenamefont {Osellame}, \citenamefont {Spagnolo},\
  and\ \citenamefont {Sciarrino}}]{brod2019p}%
  \BibitemOpen
  \bibfield  {author} {\bibinfo {author} {\bibfnamefont {D.~J.}\ \bibnamefont
  {Brod}}, \bibinfo {author} {\bibfnamefont {E.~F.}\ \bibnamefont
  {Galv{\~a}o}}, \bibinfo {author} {\bibfnamefont {A.}~\bibnamefont {Crespi}},
  \bibinfo {author} {\bibfnamefont {R.}~\bibnamefont {Osellame}}, \bibinfo
  {author} {\bibfnamefont {N.}~\bibnamefont {Spagnolo}},\ and\ \bibinfo
  {author} {\bibfnamefont {F.}~\bibnamefont {Sciarrino}},\ }\bibfield  {title}
  {\bibinfo {title} {Photonic implementation of boson sampling: a review},\
  }\href {https://doi.org/10.1117/1.AP.1.3.034001} {\bibfield  {journal}
  {\bibinfo  {journal} {Adv. Photon.}\ }\textbf {\bibinfo {volume} {1}},\
  \bibinfo {pages} {034001} (\bibinfo {year} {2019}{\natexlab{a}})}\BibitemShut
  {NoStop}%
\bibitem [{\citenamefont {Zhong}\ \emph {et~al.}(2020)\citenamefont {Zhong},
  \citenamefont {Wang}, \citenamefont {Deng}, \citenamefont {Chen},
  \citenamefont {Peng}, \citenamefont {Luo}, \citenamefont {Qin}, \citenamefont
  {Wu}, \citenamefont {Ding}, \citenamefont {Hu} \emph {et~al.}}]{zhong2020}%
  \BibitemOpen
  \bibfield  {author} {\bibinfo {author} {\bibfnamefont {H.-S.}\ \bibnamefont
  {Zhong}}, \bibinfo {author} {\bibfnamefont {H.}~\bibnamefont {Wang}},
  \bibinfo {author} {\bibfnamefont {Y.-H.}\ \bibnamefont {Deng}}, \bibinfo
  {author} {\bibfnamefont {M.-C.}\ \bibnamefont {Chen}}, \bibinfo {author}
  {\bibfnamefont {L.-C.}\ \bibnamefont {Peng}}, \bibinfo {author}
  {\bibfnamefont {Y.-H.}\ \bibnamefont {Luo}}, \bibinfo {author} {\bibfnamefont
  {J.}~\bibnamefont {Qin}}, \bibinfo {author} {\bibfnamefont {D.}~\bibnamefont
  {Wu}}, \bibinfo {author} {\bibfnamefont {X.}~\bibnamefont {Ding}}, \bibinfo
  {author} {\bibfnamefont {Y.}~\bibnamefont {Hu}}, \emph {et~al.},\ }\bibfield
  {title} {\bibinfo {title} {Quantum computational advantage using photons},\
  }\href {https://doi.org/10.1126/science.abe8770} {\bibfield  {journal}
  {\bibinfo  {journal} {Science}\ }\textbf {\bibinfo {volume} {370}},\ \bibinfo
  {pages} {1460} (\bibinfo {year} {2020})}\BibitemShut {NoStop}%
\bibitem [{\citenamefont {Zhong}\ \emph {et~al.}(2021)\citenamefont {Zhong},
  \citenamefont {Deng}, \citenamefont {Qin}, \citenamefont {Wang},
  \citenamefont {Chen}, \citenamefont {Peng}, \citenamefont {Luo},
  \citenamefont {Wu}, \citenamefont {Gong}, \citenamefont {Su} \emph
  {et~al.}}]{zhong2021}%
  \BibitemOpen
  \bibfield  {author} {\bibinfo {author} {\bibfnamefont {H.-S.}\ \bibnamefont
  {Zhong}}, \bibinfo {author} {\bibfnamefont {Y.-H.}\ \bibnamefont {Deng}},
  \bibinfo {author} {\bibfnamefont {J.}~\bibnamefont {Qin}}, \bibinfo {author}
  {\bibfnamefont {H.}~\bibnamefont {Wang}}, \bibinfo {author} {\bibfnamefont
  {M.-C.}\ \bibnamefont {Chen}}, \bibinfo {author} {\bibfnamefont {L.-C.}\
  \bibnamefont {Peng}}, \bibinfo {author} {\bibfnamefont {Y.-H.}\ \bibnamefont
  {Luo}}, \bibinfo {author} {\bibfnamefont {D.}~\bibnamefont {Wu}}, \bibinfo
  {author} {\bibfnamefont {S.-Q.}\ \bibnamefont {Gong}}, \bibinfo {author}
  {\bibfnamefont {H.}~\bibnamefont {Su}}, \emph {et~al.},\ }\bibfield  {title}
  {\bibinfo {title} {Phase-programmable gaussian boson sampling using
  stimulated squeezed light},\ }\href
  {https://doi.org/10.1103/physrevlett.127.180502} {\bibfield  {journal}
  {\bibinfo  {journal} {Phys. Rev. Lett.}\ }\textbf {\bibinfo {volume} {127}},\
  \bibinfo {pages} {180502} (\bibinfo {year} {2021})}\BibitemShut {NoStop}%
\bibitem [{\citenamefont {Spring}\ \emph {et~al.}(2017)\citenamefont {Spring},
  \citenamefont {Mennea}, \citenamefont {Metcalf}, \citenamefont {Humphreys},
  \citenamefont {Gates}, \citenamefont {Rogers}, \citenamefont {S\"{o}ller},
  \citenamefont {Smith}, \citenamefont {Kolthammer}, \citenamefont {Smith},\
  and\ \citenamefont {Walmsley}}]{Spring2017}%
  \BibitemOpen
  \bibfield  {author} {\bibinfo {author} {\bibfnamefont {J.~B.}\ \bibnamefont
  {Spring}}, \bibinfo {author} {\bibfnamefont {P.~L.}\ \bibnamefont {Mennea}},
  \bibinfo {author} {\bibfnamefont {B.~J.}\ \bibnamefont {Metcalf}}, \bibinfo
  {author} {\bibfnamefont {P.~C.}\ \bibnamefont {Humphreys}}, \bibinfo {author}
  {\bibfnamefont {J.~C.}\ \bibnamefont {Gates}}, \bibinfo {author}
  {\bibfnamefont {H.~L.}\ \bibnamefont {Rogers}}, \bibinfo {author}
  {\bibfnamefont {C.}~\bibnamefont {S\"{o}ller}}, \bibinfo {author}
  {\bibfnamefont {B.~J.}\ \bibnamefont {Smith}}, \bibinfo {author}
  {\bibfnamefont {W.~S.}\ \bibnamefont {Kolthammer}}, \bibinfo {author}
  {\bibfnamefont {P.~G.~R.}\ \bibnamefont {Smith}},\ and\ \bibinfo {author}
  {\bibfnamefont {I.~A.}\ \bibnamefont {Walmsley}},\ }\bibfield  {title}
  {\bibinfo {title} {Chip-based array of near-identical, pure, heralded
  single-photon sources},\ }\href {https://doi.org/10.1364/OPTICA.4.000090}
  {\bibfield  {journal} {\bibinfo  {journal} {Optica}\ }\textbf {\bibinfo
  {volume} {4}},\ \bibinfo {pages} {90} (\bibinfo {year} {2017})}\BibitemShut
  {NoStop}%
\bibitem [{\citenamefont {Kaneda}\ and\ \citenamefont
  {Kwiat}(2019)}]{Kaneda2019}%
  \BibitemOpen
  \bibfield  {author} {\bibinfo {author} {\bibfnamefont {F.}~\bibnamefont
  {Kaneda}}\ and\ \bibinfo {author} {\bibfnamefont {P.~G.}\ \bibnamefont
  {Kwiat}},\ }\bibfield  {title} {\bibinfo {title} {High-efficiency
  single-photon generation via large-scale active time multiplexing},\ }\href
  {https://doi.org/10.1126/sciadv.aaw8586} {\bibfield  {journal} {\bibinfo
  {journal} {Sci. Adv.}\ }\textbf {\bibinfo {volume} {5}},\ \bibinfo {pages}
  {eaaw8586} (\bibinfo {year} {2019})}\BibitemShut {NoStop}%
\bibitem [{\citenamefont {Senellart}\ \emph {et~al.}(2017)\citenamefont
  {Senellart}, \citenamefont {Solomon},\ and\ \citenamefont
  {White}}]{senellart2017}%
  \BibitemOpen
  \bibfield  {author} {\bibinfo {author} {\bibfnamefont {P.}~\bibnamefont
  {Senellart}}, \bibinfo {author} {\bibfnamefont {G.}~\bibnamefont {Solomon}},\
  and\ \bibinfo {author} {\bibfnamefont {A.}~\bibnamefont {White}},\ }\bibfield
   {title} {\bibinfo {title} {High-performance semiconductor quantum-dot
  single-photon sources},\ }\href {https://doi.org/10.1038/nnano.2017.218}
  {\bibfield  {journal} {\bibinfo  {journal} {Nat. Nanotechnol.}\ }\textbf
  {\bibinfo {volume} {12}},\ \bibinfo {pages} {1026} (\bibinfo {year}
  {2017})}\BibitemShut {NoStop}%
\bibitem [{\citenamefont {Wang}\ \emph {et~al.}(2019)\citenamefont {Wang},
  \citenamefont {He}, \citenamefont {Chung}, \citenamefont {Hu}, \citenamefont
  {Yu}, \citenamefont {Chen}, \citenamefont {Ding}, \citenamefont {Chen},
  \citenamefont {Qin}, \citenamefont {Yang}, \citenamefont {Liu}, \citenamefont
  {Duan}, \citenamefont {Li}, \citenamefont {Gerhardt}, \citenamefont
  {Winkler}, \citenamefont {Jurkat}, \citenamefont {Wang}, \citenamefont
  {Gregersen}, \citenamefont {Huo}, \citenamefont {Dai}, \citenamefont {Yu},
  \citenamefont {H{\"o}fling}, \citenamefont {Lu},\ and\ \citenamefont
  {Pan}}]{Wang2019}%
  \BibitemOpen
  \bibfield  {author} {\bibinfo {author} {\bibfnamefont {H.}~\bibnamefont
  {Wang}}, \bibinfo {author} {\bibfnamefont {Y.-M.}\ \bibnamefont {He}},
  \bibinfo {author} {\bibfnamefont {T.-H.}\ \bibnamefont {Chung}}, \bibinfo
  {author} {\bibfnamefont {H.}~\bibnamefont {Hu}}, \bibinfo {author}
  {\bibfnamefont {Y.}~\bibnamefont {Yu}}, \bibinfo {author} {\bibfnamefont
  {S.}~\bibnamefont {Chen}}, \bibinfo {author} {\bibfnamefont {X.}~\bibnamefont
  {Ding}}, \bibinfo {author} {\bibfnamefont {M.-C.}\ \bibnamefont {Chen}},
  \bibinfo {author} {\bibfnamefont {J.}~\bibnamefont {Qin}}, \bibinfo {author}
  {\bibfnamefont {X.}~\bibnamefont {Yang}}, \bibinfo {author} {\bibfnamefont
  {R.-Z.}\ \bibnamefont {Liu}}, \bibinfo {author} {\bibfnamefont {Z.-C.}\
  \bibnamefont {Duan}}, \bibinfo {author} {\bibfnamefont {J.-P.}\ \bibnamefont
  {Li}}, \bibinfo {author} {\bibfnamefont {S.}~\bibnamefont {Gerhardt}},
  \bibinfo {author} {\bibfnamefont {K.}~\bibnamefont {Winkler}}, \bibinfo
  {author} {\bibfnamefont {J.}~\bibnamefont {Jurkat}}, \bibinfo {author}
  {\bibfnamefont {L.-J.}\ \bibnamefont {Wang}}, \bibinfo {author}
  {\bibfnamefont {N.}~\bibnamefont {Gregersen}}, \bibinfo {author}
  {\bibfnamefont {Y.-H.}\ \bibnamefont {Huo}}, \bibinfo {author} {\bibfnamefont
  {Q.}~\bibnamefont {Dai}}, \bibinfo {author} {\bibfnamefont {S.}~\bibnamefont
  {Yu}}, \bibinfo {author} {\bibfnamefont {S.}~\bibnamefont {H{\"o}fling}},
  \bibinfo {author} {\bibfnamefont {C.-Y.}\ \bibnamefont {Lu}},\ and\ \bibinfo
  {author} {\bibfnamefont {J.-W.}\ \bibnamefont {Pan}},\ }\bibfield  {title}
  {\bibinfo {title} {Towards optimal single-photon sources from polarized
  microcavities},\ }\href {https://doi.org/10.1038/s41566-019-0494-3}
  {\bibfield  {journal} {\bibinfo  {journal} {Nat. Photon.}\ }\textbf {\bibinfo
  {volume} {13}},\ \bibinfo {pages} {770} (\bibinfo {year} {2019})}\BibitemShut
  {NoStop}%
\bibitem [{\citenamefont {Tomm}\ \emph {et~al.}(2021)\citenamefont {Tomm},
  \citenamefont {Javadi}, \citenamefont {Antoniadis}, \citenamefont {Najer},
  \citenamefont {Lobl}, \citenamefont {Korsch}, \citenamefont {Schott},
  \citenamefont {Valentin}, \citenamefont {Wieck}, \citenamefont {Ludwig},\
  and\ \citenamefont {Warburton}}]{Tomm2021}%
  \BibitemOpen
  \bibfield  {author} {\bibinfo {author} {\bibfnamefont {N.}~\bibnamefont
  {Tomm}}, \bibinfo {author} {\bibfnamefont {A.}~\bibnamefont {Javadi}},
  \bibinfo {author} {\bibfnamefont {N.~O.}\ \bibnamefont {Antoniadis}},
  \bibinfo {author} {\bibfnamefont {D.}~\bibnamefont {Najer}}, \bibinfo
  {author} {\bibfnamefont {M.~C.}\ \bibnamefont {Lobl}}, \bibinfo {author}
  {\bibfnamefont {A.~R.}\ \bibnamefont {Korsch}}, \bibinfo {author}
  {\bibfnamefont {R.}~\bibnamefont {Schott}}, \bibinfo {author} {\bibfnamefont
  {S.~R.}\ \bibnamefont {Valentin}}, \bibinfo {author} {\bibfnamefont {A.~D.}\
  \bibnamefont {Wieck}}, \bibinfo {author} {\bibfnamefont {A.}~\bibnamefont
  {Ludwig}},\ and\ \bibinfo {author} {\bibfnamefont {R.~J.}\ \bibnamefont
  {Warburton}},\ }\bibfield  {title} {\bibinfo {title} {A bright and fast
  source of coherent single photons},\ }\bibfield  {journal} {\bibinfo
  {journal} {Nat. Nanotechnol.}\ }\href
  {https://doi.org/10.1038/s41565-020-00831-x} {10.1038/s41565-020-00831-x}
  (\bibinfo {year} {2021})\BibitemShut {NoStop}%
\bibitem [{\citenamefont {Thomas}\ \emph {et~al.}(2021)\citenamefont {Thomas},
  \citenamefont {Billard}, \citenamefont {Coste}, \citenamefont {Wein},
  \citenamefont {Priya}, \citenamefont {Ollivier}, \citenamefont {Krebs},
  \citenamefont {Taza\"{\i}rt}, \citenamefont {Harouri}, \citenamefont
  {Lemaitre}, \citenamefont {Sagnes}, \citenamefont {Anton}, \citenamefont
  {Lanco}, \citenamefont {Somaschi}, \citenamefont {Loredo},\ and\
  \citenamefont {Senellart}}]{Thomas2021}%
  \BibitemOpen
  \bibfield  {author} {\bibinfo {author} {\bibfnamefont {S.~E.}\ \bibnamefont
  {Thomas}}, \bibinfo {author} {\bibfnamefont {M.}~\bibnamefont {Billard}},
  \bibinfo {author} {\bibfnamefont {N.}~\bibnamefont {Coste}}, \bibinfo
  {author} {\bibfnamefont {S.~C.}\ \bibnamefont {Wein}}, \bibinfo {author}
  {\bibnamefont {Priya}}, \bibinfo {author} {\bibfnamefont {H.}~\bibnamefont
  {Ollivier}}, \bibinfo {author} {\bibfnamefont {O.}~\bibnamefont {Krebs}},
  \bibinfo {author} {\bibfnamefont {L.}~\bibnamefont {Taza\"{\i}rt}}, \bibinfo
  {author} {\bibfnamefont {A.}~\bibnamefont {Harouri}}, \bibinfo {author}
  {\bibfnamefont {A.}~\bibnamefont {Lemaitre}}, \bibinfo {author}
  {\bibfnamefont {I.}~\bibnamefont {Sagnes}}, \bibinfo {author} {\bibfnamefont
  {C.}~\bibnamefont {Anton}}, \bibinfo {author} {\bibfnamefont
  {L.}~\bibnamefont {Lanco}}, \bibinfo {author} {\bibfnamefont
  {N.}~\bibnamefont {Somaschi}}, \bibinfo {author} {\bibfnamefont {J.~C.}\
  \bibnamefont {Loredo}},\ and\ \bibinfo {author} {\bibfnamefont
  {P.}~\bibnamefont {Senellart}},\ }\bibfield  {title} {\bibinfo {title}
  {Bright polarized single-photon source based on a linear dipole},\ }\href
  {https://doi.org/10.1103/PhysRevLett.126.233601} {\bibfield  {journal}
  {\bibinfo  {journal} {Phys. Rev. Lett.}\ }\textbf {\bibinfo {volume} {126}},\
  \bibinfo {pages} {233601} (\bibinfo {year} {2021})}\BibitemShut {NoStop}%
\bibitem [{\citenamefont {Hong}\ \emph {et~al.}(1987)\citenamefont {Hong},
  \citenamefont {Ou},\ and\ \citenamefont {Mandel}}]{HOM}%
  \BibitemOpen
  \bibfield  {author} {\bibinfo {author} {\bibfnamefont {C.~K.}\ \bibnamefont
  {Hong}}, \bibinfo {author} {\bibfnamefont {Z.~Y.}\ \bibnamefont {Ou}},\ and\
  \bibinfo {author} {\bibfnamefont {L.}~\bibnamefont {Mandel}},\ }\bibfield
  {title} {\bibinfo {title} {Measurement of subpicosecond time intervals
  between two photons by interference},\ }\href
  {https://doi.org/10.1103/PhysRevLett.59.2044} {\bibfield  {journal} {\bibinfo
   {journal} {Phys. Rev. Lett.}\ }\textbf {\bibinfo {volume} {59}},\ \bibinfo
  {pages} {2044} (\bibinfo {year} {1987})}\BibitemShut {NoStop}%
\bibitem [{\citenamefont {Menssen}\ \emph {et~al.}(2017)\citenamefont
  {Menssen}, \citenamefont {Jones}, \citenamefont {Metcalf}, \citenamefont
  {Tichy}, \citenamefont {Barz}, \citenamefont {Kolthammer},\ and\
  \citenamefont {Walmsley}}]{menssen2017}%
  \BibitemOpen
  \bibfield  {author} {\bibinfo {author} {\bibfnamefont {A.~J.}\ \bibnamefont
  {Menssen}}, \bibinfo {author} {\bibfnamefont {A.~E.}\ \bibnamefont {Jones}},
  \bibinfo {author} {\bibfnamefont {B.~J.}\ \bibnamefont {Metcalf}}, \bibinfo
  {author} {\bibfnamefont {M.~C.}\ \bibnamefont {Tichy}}, \bibinfo {author}
  {\bibfnamefont {S.}~\bibnamefont {Barz}}, \bibinfo {author} {\bibfnamefont
  {W.~S.}\ \bibnamefont {Kolthammer}},\ and\ \bibinfo {author} {\bibfnamefont
  {I.~A.}\ \bibnamefont {Walmsley}},\ }\bibfield  {title} {\bibinfo {title}
  {Distinguishability and many-particle interference},\ }\href
  {https://doi.org/10.1103/physrevlett.118.153603} {\bibfield  {journal}
  {\bibinfo  {journal} {Phys. Rev. Lett.}\ }\textbf {\bibinfo {volume} {118}},\
  \bibinfo {pages} {153603} (\bibinfo {year} {2017})}\BibitemShut {NoStop}%
\bibitem [{\citenamefont {Fischer}\ \emph {et~al.}(2018)\citenamefont
  {Fischer}, \citenamefont {Trivedi},\ and\ \citenamefont
  {Lukin}}]{Fischer2018}%
  \BibitemOpen
  \bibfield  {author} {\bibinfo {author} {\bibfnamefont {K.~A.}\ \bibnamefont
  {Fischer}}, \bibinfo {author} {\bibfnamefont {R.}~\bibnamefont {Trivedi}},\
  and\ \bibinfo {author} {\bibfnamefont {D.}~\bibnamefont {Lukin}},\ }\bibfield
   {title} {\bibinfo {title} {Particle emission from open quantum systems},\
  }\href {https://doi.org/10.1103/PhysRevA.98.023853} {\bibfield  {journal}
  {\bibinfo  {journal} {Phys. Rev. A}\ }\textbf {\bibinfo {volume} {98}},\
  \bibinfo {pages} {023853} (\bibinfo {year} {2018})}\BibitemShut {NoStop}%
\bibitem [{\citenamefont {Trivedi}\ \emph {et~al.}(2020)\citenamefont
  {Trivedi}, \citenamefont {Fischer}, \citenamefont {Vučković},\ and\
  \citenamefont {Müller}}]{Trivedi2020}%
  \BibitemOpen
  \bibfield  {author} {\bibinfo {author} {\bibfnamefont {R.}~\bibnamefont
  {Trivedi}}, \bibinfo {author} {\bibfnamefont {K.~A.}\ \bibnamefont
  {Fischer}}, \bibinfo {author} {\bibfnamefont {J.}~\bibnamefont
  {Vučković}},\ and\ \bibinfo {author} {\bibfnamefont {K.}~\bibnamefont
  {Müller}},\ }\bibfield  {title} {\bibinfo {title} {Generation of
  non-classical light using semiconductor quantum dots},\ }\href
  {https://doi.org/10.1002/qute.201900007} {\bibfield  {journal} {\bibinfo
  {journal} {Adv. Quantum Technol.}\ }\textbf {\bibinfo {volume} {3}},\
  \bibinfo {pages} {1900007} (\bibinfo {year} {2020})}\BibitemShut {NoStop}%
\bibitem [{\citenamefont {Ollivier}\ \emph {et~al.}(2021)\citenamefont
  {Ollivier}, \citenamefont {Thomas}, \citenamefont {Wein}, \citenamefont
  {de~Buy~Wenniger}, \citenamefont {Coste}, \citenamefont {Loredo},
  \citenamefont {Somaschi}, \citenamefont {Harouri}, \citenamefont {Lemaitre},
  \citenamefont {Sagnes}, \citenamefont {Lanco}, \citenamefont {Simon},
  \citenamefont {Anton}, \citenamefont {Krebs},\ and\ \citenamefont
  {Senellart}}]{Ollivier2021}%
  \BibitemOpen
  \bibfield  {author} {\bibinfo {author} {\bibfnamefont {H.}~\bibnamefont
  {Ollivier}}, \bibinfo {author} {\bibfnamefont {S.~E.}\ \bibnamefont
  {Thomas}}, \bibinfo {author} {\bibfnamefont {S.~C.}\ \bibnamefont {Wein}},
  \bibinfo {author} {\bibfnamefont {I.~M.}\ \bibnamefont {de~Buy~Wenniger}},
  \bibinfo {author} {\bibfnamefont {N.}~\bibnamefont {Coste}}, \bibinfo
  {author} {\bibfnamefont {J.~C.}\ \bibnamefont {Loredo}}, \bibinfo {author}
  {\bibfnamefont {N.}~\bibnamefont {Somaschi}}, \bibinfo {author}
  {\bibfnamefont {A.}~\bibnamefont {Harouri}}, \bibinfo {author} {\bibfnamefont
  {A.}~\bibnamefont {Lemaitre}}, \bibinfo {author} {\bibfnamefont
  {I.}~\bibnamefont {Sagnes}}, \bibinfo {author} {\bibfnamefont
  {L.}~\bibnamefont {Lanco}}, \bibinfo {author} {\bibfnamefont
  {C.}~\bibnamefont {Simon}}, \bibinfo {author} {\bibfnamefont
  {C.}~\bibnamefont {Anton}}, \bibinfo {author} {\bibfnamefont
  {O.}~\bibnamefont {Krebs}},\ and\ \bibinfo {author} {\bibfnamefont
  {P.}~\bibnamefont {Senellart}},\ }\bibfield  {title} {\bibinfo {title}
  {Hong-ou-mandel interference with imperfect single photon sources},\ }\href
  {https://doi.org/10.1103/PhysRevLett.126.063602} {\bibfield  {journal}
  {\bibinfo  {journal} {Phys. Rev. Lett.}\ }\textbf {\bibinfo {volume} {126}},\
  \bibinfo {pages} {063602} (\bibinfo {year} {2021})}\BibitemShut {NoStop}%
\bibitem [{\citenamefont {Giordani}\ \emph {et~al.}(2020)\citenamefont
  {Giordani}, \citenamefont {Brod}, \citenamefont {Esposito}, \citenamefont
  {Viggianiello}, \citenamefont {Romano}, \citenamefont {Flamini},
  \citenamefont {Carvacho}, \citenamefont {Spagnolo}, \citenamefont
  {Galv{\~a}o},\ and\ \citenamefont {Sciarrino}}]{giordani2020}%
  \BibitemOpen
  \bibfield  {author} {\bibinfo {author} {\bibfnamefont {T.}~\bibnamefont
  {Giordani}}, \bibinfo {author} {\bibfnamefont {D.~J.}\ \bibnamefont {Brod}},
  \bibinfo {author} {\bibfnamefont {C.}~\bibnamefont {Esposito}}, \bibinfo
  {author} {\bibfnamefont {N.}~\bibnamefont {Viggianiello}}, \bibinfo {author}
  {\bibfnamefont {M.}~\bibnamefont {Romano}}, \bibinfo {author} {\bibfnamefont
  {F.}~\bibnamefont {Flamini}}, \bibinfo {author} {\bibfnamefont
  {G.}~\bibnamefont {Carvacho}}, \bibinfo {author} {\bibfnamefont
  {N.}~\bibnamefont {Spagnolo}}, \bibinfo {author} {\bibfnamefont {E.~F.}\
  \bibnamefont {Galv{\~a}o}},\ and\ \bibinfo {author} {\bibfnamefont
  {F.}~\bibnamefont {Sciarrino}},\ }\bibfield  {title} {\bibinfo {title}
  {Experimental quantification of four-photon indistinguishability},\ }\href
  {https://doi.org/10.1088/1367-2630/ab7a30} {\bibfield  {journal} {\bibinfo
  {journal} {New J. Phys.}\ }\textbf {\bibinfo {volume} {22}},\ \bibinfo
  {pages} {043001} (\bibinfo {year} {2020})}\BibitemShut {NoStop}%
\bibitem [{\citenamefont {Brod}\ \emph
  {et~al.}(2019{\natexlab{b}})\citenamefont {Brod}, \citenamefont {Galv{\~a}o},
  \citenamefont {Viggianiello}, \citenamefont {Flamini}, \citenamefont
  {Spagnolo},\ and\ \citenamefont {Sciarrino}}]{brod2019w}%
  \BibitemOpen
  \bibfield  {author} {\bibinfo {author} {\bibfnamefont {D.~J.}\ \bibnamefont
  {Brod}}, \bibinfo {author} {\bibfnamefont {E.~F.}\ \bibnamefont
  {Galv{\~a}o}}, \bibinfo {author} {\bibfnamefont {N.}~\bibnamefont
  {Viggianiello}}, \bibinfo {author} {\bibfnamefont {F.}~\bibnamefont
  {Flamini}}, \bibinfo {author} {\bibfnamefont {N.}~\bibnamefont {Spagnolo}},\
  and\ \bibinfo {author} {\bibfnamefont {F.}~\bibnamefont {Sciarrino}},\
  }\bibfield  {title} {\bibinfo {title} {Witnessing genuine multiphoton
  indistinguishability},\ }\href
  {https://doi.org/10.1103/physrevlett.122.063602} {\bibfield  {journal}
  {\bibinfo  {journal} {Phys. Rev. Lett.}\ }\textbf {\bibinfo {volume} {122}},\
  \bibinfo {pages} {063602} (\bibinfo {year} {2019}{\natexlab{b}})}\BibitemShut
  {NoStop}%
\bibitem [{\citenamefont {Tichy}\ \emph {et~al.}(2010)\citenamefont {Tichy},
  \citenamefont {Tiersch}, \citenamefont {de~Melo}, \citenamefont {Mintert},\
  and\ \citenamefont {Buchleitner}}]{tichy2010}%
  \BibitemOpen
  \bibfield  {author} {\bibinfo {author} {\bibfnamefont {M.~C.}\ \bibnamefont
  {Tichy}}, \bibinfo {author} {\bibfnamefont {M.}~\bibnamefont {Tiersch}},
  \bibinfo {author} {\bibfnamefont {F.}~\bibnamefont {de~Melo}}, \bibinfo
  {author} {\bibfnamefont {F.}~\bibnamefont {Mintert}},\ and\ \bibinfo {author}
  {\bibfnamefont {A.}~\bibnamefont {Buchleitner}},\ }\bibfield  {title}
  {\bibinfo {title} {Zero-transmission law for multiport beam splitters},\
  }\href {https://doi.org/10.1103/physrevlett.104.220405} {\bibfield  {journal}
  {\bibinfo  {journal} {Phys. Rev. Lett.}\ }\textbf {\bibinfo {volume} {104}},\
  \bibinfo {pages} {220405} (\bibinfo {year} {2010})}\BibitemShut {NoStop}%
\bibitem [{\citenamefont {Crespi}(2015)}]{crespi2015}%
  \BibitemOpen
  \bibfield  {author} {\bibinfo {author} {\bibfnamefont {A.}~\bibnamefont
  {Crespi}},\ }\bibfield  {title} {\bibinfo {title} {Suppression laws for
  multiparticle interference in sylvester interferometers},\ }\href
  {https://doi.org/10.1103/physreva.91.013811} {\bibfield  {journal} {\bibinfo
  {journal} {Phys. Rev. A}\ }\textbf {\bibinfo {volume} {91}},\ \bibinfo
  {pages} {013811} (\bibinfo {year} {2015})}\BibitemShut {NoStop}%
\bibitem [{\citenamefont {Dittel}\ \emph {et~al.}(2017)\citenamefont {Dittel},
  \citenamefont {Keil},\ and\ \citenamefont {Weihs}}]{dittel2017}%
  \BibitemOpen
  \bibfield  {author} {\bibinfo {author} {\bibfnamefont {C.}~\bibnamefont
  {Dittel}}, \bibinfo {author} {\bibfnamefont {R.}~\bibnamefont {Keil}},\ and\
  \bibinfo {author} {\bibfnamefont {G.}~\bibnamefont {Weihs}},\ }\bibfield
  {title} {\bibinfo {title} {Many-body quantum interference on hypercubes},\
  }\href {https://doi.org/10.1088/2058-9565/aa540c} {\bibfield  {journal}
  {\bibinfo  {journal} {Quantum Sci. Technol.}\ }\textbf {\bibinfo {volume}
  {2}},\ \bibinfo {pages} {015003} (\bibinfo {year} {2017})}\BibitemShut
  {NoStop}%
\bibitem [{\citenamefont {Dittel}\ \emph {et~al.}(2018)\citenamefont {Dittel},
  \citenamefont {Dufour}, \citenamefont {Walschaers}, \citenamefont {Weihs},
  \citenamefont {Buchleitner},\ and\ \citenamefont {Keil}}]{dittel2018}%
  \BibitemOpen
  \bibfield  {author} {\bibinfo {author} {\bibfnamefont {C.}~\bibnamefont
  {Dittel}}, \bibinfo {author} {\bibfnamefont {G.}~\bibnamefont {Dufour}},
  \bibinfo {author} {\bibfnamefont {M.}~\bibnamefont {Walschaers}}, \bibinfo
  {author} {\bibfnamefont {G.}~\bibnamefont {Weihs}}, \bibinfo {author}
  {\bibfnamefont {A.}~\bibnamefont {Buchleitner}},\ and\ \bibinfo {author}
  {\bibfnamefont {R.}~\bibnamefont {Keil}},\ }\bibfield  {title} {\bibinfo
  {title} {Totally destructive many-particle interference},\ }\href
  {https://doi.org/10.1103/physrevlett.120.240404} {\bibfield  {journal}
  {\bibinfo  {journal} {Phys. Rev. Lett.}\ }\textbf {\bibinfo {volume} {120}},\
  \bibinfo {pages} {240404} (\bibinfo {year} {2018})}\BibitemShut {NoStop}%
\bibitem [{\citenamefont {Viggianiello}\ \emph
  {et~al.}(2018{\natexlab{a}})\citenamefont {Viggianiello}, \citenamefont
  {Flamini}, \citenamefont {Innocenti}, \citenamefont {Cozzolino},
  \citenamefont {Bentivegna}, \citenamefont {Spagnolo}, \citenamefont {Crespi},
  \citenamefont {Brod}, \citenamefont {Galv{\~a}o}, \citenamefont {Osellame}
  \emph {et~al.}}]{viggianiello2018}%
  \BibitemOpen
  \bibfield  {author} {\bibinfo {author} {\bibfnamefont {N.}~\bibnamefont
  {Viggianiello}}, \bibinfo {author} {\bibfnamefont {F.}~\bibnamefont
  {Flamini}}, \bibinfo {author} {\bibfnamefont {L.}~\bibnamefont {Innocenti}},
  \bibinfo {author} {\bibfnamefont {D.}~\bibnamefont {Cozzolino}}, \bibinfo
  {author} {\bibfnamefont {M.}~\bibnamefont {Bentivegna}}, \bibinfo {author}
  {\bibfnamefont {N.}~\bibnamefont {Spagnolo}}, \bibinfo {author}
  {\bibfnamefont {A.}~\bibnamefont {Crespi}}, \bibinfo {author} {\bibfnamefont
  {D.~J.}\ \bibnamefont {Brod}}, \bibinfo {author} {\bibfnamefont {E.~F.}\
  \bibnamefont {Galv{\~a}o}}, \bibinfo {author} {\bibfnamefont
  {R.}~\bibnamefont {Osellame}}, \emph {et~al.},\ }\bibfield  {title} {\bibinfo
  {title} {Experimental generalized quantum suppression law in sylvester
  interferometers},\ }\href {https://doi.org/10.1088/1367-2630/aaad92}
  {\bibfield  {journal} {\bibinfo  {journal} {New J. Phys.}\ }\textbf {\bibinfo
  {volume} {20}},\ \bibinfo {pages} {033017} (\bibinfo {year}
  {2018}{\natexlab{a}})}\BibitemShut {NoStop}%
\bibitem [{\citenamefont {Crespi}\ \emph {et~al.}(2016)\citenamefont {Crespi},
  \citenamefont {Osellame}, \citenamefont {Ramponi}, \citenamefont
  {Bentivegna}, \citenamefont {Flamini}, \citenamefont {Spagnolo},
  \citenamefont {Viggianiello}, \citenamefont {Innocenti}, \citenamefont
  {Mataloni},\ and\ \citenamefont {Sciarrino}}]{crespi2016}%
  \BibitemOpen
  \bibfield  {author} {\bibinfo {author} {\bibfnamefont {A.}~\bibnamefont
  {Crespi}}, \bibinfo {author} {\bibfnamefont {R.}~\bibnamefont {Osellame}},
  \bibinfo {author} {\bibfnamefont {R.}~\bibnamefont {Ramponi}}, \bibinfo
  {author} {\bibfnamefont {M.}~\bibnamefont {Bentivegna}}, \bibinfo {author}
  {\bibfnamefont {F.}~\bibnamefont {Flamini}}, \bibinfo {author} {\bibfnamefont
  {N.}~\bibnamefont {Spagnolo}}, \bibinfo {author} {\bibfnamefont
  {N.}~\bibnamefont {Viggianiello}}, \bibinfo {author} {\bibfnamefont
  {L.}~\bibnamefont {Innocenti}}, \bibinfo {author} {\bibfnamefont
  {P.}~\bibnamefont {Mataloni}},\ and\ \bibinfo {author} {\bibfnamefont
  {F.}~\bibnamefont {Sciarrino}},\ }\bibfield  {title} {\bibinfo {title}
  {Suppression law of quantum states in a 3d photonic fast fourier transform
  chip},\ }\href {https://doi.org/10.1038/ncomms10469} {\bibfield  {journal}
  {\bibinfo  {journal} {Nat. Commun.}\ }\textbf {\bibinfo {volume} {7}},\
  \bibinfo {pages} {1} (\bibinfo {year} {2016})}\BibitemShut {NoStop}%
\bibitem [{\citenamefont {Tichy}\ \emph {et~al.}(2014)\citenamefont {Tichy},
  \citenamefont {Mayer}, \citenamefont {Buchleitner},\ and\ \citenamefont
  {M{\o}lmer}}]{tichy2014}%
  \BibitemOpen
  \bibfield  {author} {\bibinfo {author} {\bibfnamefont {M.~C.}\ \bibnamefont
  {Tichy}}, \bibinfo {author} {\bibfnamefont {K.}~\bibnamefont {Mayer}},
  \bibinfo {author} {\bibfnamefont {A.}~\bibnamefont {Buchleitner}},\ and\
  \bibinfo {author} {\bibfnamefont {K.}~\bibnamefont {M{\o}lmer}},\ }\bibfield
  {title} {\bibinfo {title} {Stringent and efficient assessment of
  boson-sampling devices},\ }\href
  {https://doi.org/10.1103/physrevlett.113.020502} {\bibfield  {journal}
  {\bibinfo  {journal} {Phys. Rev. Lett.}\ }\textbf {\bibinfo {volume} {113}},\
  \bibinfo {pages} {020502} (\bibinfo {year} {2014})}\BibitemShut {NoStop}%
\bibitem [{\citenamefont {Viggianiello}\ \emph
  {et~al.}(2018{\natexlab{b}})\citenamefont {Viggianiello}, \citenamefont
  {Flamini}, \citenamefont {Bentivegna}, \citenamefont {Spagnolo},
  \citenamefont {Crespi}, \citenamefont {Brod}, \citenamefont {Galv{\~a}o},
  \citenamefont {Osellame},\ and\ \citenamefont
  {Sciarrino}}]{viggianiello2018opt}%
  \BibitemOpen
  \bibfield  {author} {\bibinfo {author} {\bibfnamefont {N.}~\bibnamefont
  {Viggianiello}}, \bibinfo {author} {\bibfnamefont {F.}~\bibnamefont
  {Flamini}}, \bibinfo {author} {\bibfnamefont {M.}~\bibnamefont {Bentivegna}},
  \bibinfo {author} {\bibfnamefont {N.}~\bibnamefont {Spagnolo}}, \bibinfo
  {author} {\bibfnamefont {A.}~\bibnamefont {Crespi}}, \bibinfo {author}
  {\bibfnamefont {D.~J.}\ \bibnamefont {Brod}}, \bibinfo {author}
  {\bibfnamefont {E.~F.}\ \bibnamefont {Galv{\~a}o}}, \bibinfo {author}
  {\bibfnamefont {R.}~\bibnamefont {Osellame}},\ and\ \bibinfo {author}
  {\bibfnamefont {F.}~\bibnamefont {Sciarrino}},\ }\bibfield  {title} {\bibinfo
  {title} {Optimal photonic indistinguishability tests in multimode networks},\
  }\href {https://doi.org/10.1016/j.scib.2018.10.009} {\bibfield  {journal}
  {\bibinfo  {journal} {Sci. Bull.}\ }\textbf {\bibinfo {volume} {63}},\
  \bibinfo {pages} {1470} (\bibinfo {year} {2018}{\natexlab{b}})}\BibitemShut
  {NoStop}%
\bibitem [{\citenamefont {M{\"u}nzberg}\ \emph {et~al.}(2021)\citenamefont
  {M{\"u}nzberg}, \citenamefont {Dittel}, \citenamefont {Lebugle},
  \citenamefont {Buchleitner}, \citenamefont {Szameit}, \citenamefont {Weihs},\
  and\ \citenamefont {Keil}}]{munzberg2021}%
  \BibitemOpen
  \bibfield  {author} {\bibinfo {author} {\bibfnamefont {J.}~\bibnamefont
  {M{\"u}nzberg}}, \bibinfo {author} {\bibfnamefont {C.}~\bibnamefont
  {Dittel}}, \bibinfo {author} {\bibfnamefont {M.}~\bibnamefont {Lebugle}},
  \bibinfo {author} {\bibfnamefont {A.}~\bibnamefont {Buchleitner}}, \bibinfo
  {author} {\bibfnamefont {A.}~\bibnamefont {Szameit}}, \bibinfo {author}
  {\bibfnamefont {G.}~\bibnamefont {Weihs}},\ and\ \bibinfo {author}
  {\bibfnamefont {R.}~\bibnamefont {Keil}},\ }\bibfield  {title} {\bibinfo
  {title} {Symmetry allows for distinguishability in totally destructive
  many-particle interference},\ }\href
  {https://doi.org/10.1103/prxquantum.2.020326} {\bibfield  {journal} {\bibinfo
   {journal} {PRX Quantum}\ }\textbf {\bibinfo {volume} {2}},\ \bibinfo {pages}
  {020326} (\bibinfo {year} {2021})}\BibitemShut {NoStop}%
\bibitem [{\citenamefont {van~der Meer}\ \emph {et~al.}(2021)\citenamefont
  {van~der Meer}, \citenamefont {Hooijschuur}, \citenamefont {Somhorst},
  \citenamefont {Venderbosch}, \citenamefont {de~Goede}, \citenamefont
  {Kassenberg}, \citenamefont {Snijders}, \citenamefont {Taballione},
  \citenamefont {Epping}, \citenamefont {Vlekkert} \emph
  {et~al.}}]{vandermeer2021}%
  \BibitemOpen
  \bibfield  {author} {\bibinfo {author} {\bibfnamefont {R.}~\bibnamefont
  {van~der Meer}}, \bibinfo {author} {\bibfnamefont {P.}~\bibnamefont
  {Hooijschuur}}, \bibinfo {author} {\bibfnamefont {F.~H.}\ \bibnamefont
  {Somhorst}}, \bibinfo {author} {\bibfnamefont {P.}~\bibnamefont
  {Venderbosch}}, \bibinfo {author} {\bibfnamefont {M.}~\bibnamefont
  {de~Goede}}, \bibinfo {author} {\bibfnamefont {B.}~\bibnamefont
  {Kassenberg}}, \bibinfo {author} {\bibfnamefont {H.}~\bibnamefont
  {Snijders}}, \bibinfo {author} {\bibfnamefont {C.}~\bibnamefont
  {Taballione}}, \bibinfo {author} {\bibfnamefont {J.}~\bibnamefont {Epping}},
  \bibinfo {author} {\bibfnamefont {H.~v.~d.}\ \bibnamefont {Vlekkert}}, \emph
  {et~al.},\ }\bibfield  {title} {\bibinfo {title} {Experimental demonstration
  of an efficient, semi-device-independent photonic indistinguishability
  witness},\ }\href {https://arxiv.org/abs/2112.00067} {\bibfield  {journal}
  {\bibinfo  {journal} {arXiv preprint arXiv:2112.00067}\ } (\bibinfo {year}
  {2021})}\BibitemShut {NoStop}%
\bibitem [{\citenamefont {Somaschi}\ \emph {et~al.}(2016)\citenamefont
  {Somaschi}, \citenamefont {Giesz}, \citenamefont {De~Santis}, \citenamefont
  {Loredo}, \citenamefont {Almeida}, \citenamefont {Hornecker}, \citenamefont
  {Portalupi}, \citenamefont {Grange}, \citenamefont {Ant{\'o}n}, \citenamefont
  {Demory}, \citenamefont {G{\'o}mez}, \citenamefont {Sagnes}, \citenamefont
  {Lanzillotti-Kimura}, \citenamefont {Lema{\'i}tre}, \citenamefont {Auffeves},
  \citenamefont {White}, \citenamefont {Lanco},\ and\ \citenamefont
  {Senellart}}]{Somaschi2016}%
  \BibitemOpen
  \bibfield  {author} {\bibinfo {author} {\bibfnamefont {N.}~\bibnamefont
  {Somaschi}}, \bibinfo {author} {\bibfnamefont {V.}~\bibnamefont {Giesz}},
  \bibinfo {author} {\bibfnamefont {L.}~\bibnamefont {De~Santis}}, \bibinfo
  {author} {\bibfnamefont {J.~C.}\ \bibnamefont {Loredo}}, \bibinfo {author}
  {\bibfnamefont {M.~P.}\ \bibnamefont {Almeida}}, \bibinfo {author}
  {\bibfnamefont {G.}~\bibnamefont {Hornecker}}, \bibinfo {author}
  {\bibfnamefont {S.~L.}\ \bibnamefont {Portalupi}}, \bibinfo {author}
  {\bibfnamefont {T.}~\bibnamefont {Grange}}, \bibinfo {author} {\bibfnamefont
  {C.}~\bibnamefont {Ant{\'o}n}}, \bibinfo {author} {\bibfnamefont
  {J.}~\bibnamefont {Demory}}, \bibinfo {author} {\bibfnamefont
  {C.}~\bibnamefont {G{\'o}mez}}, \bibinfo {author} {\bibfnamefont
  {I.}~\bibnamefont {Sagnes}}, \bibinfo {author} {\bibfnamefont {N.~D.}\
  \bibnamefont {Lanzillotti-Kimura}}, \bibinfo {author} {\bibfnamefont
  {A.}~\bibnamefont {Lema{\'i}tre}}, \bibinfo {author} {\bibfnamefont
  {A.}~\bibnamefont {Auffeves}}, \bibinfo {author} {\bibfnamefont {A.~G.}\
  \bibnamefont {White}}, \bibinfo {author} {\bibfnamefont {L.}~\bibnamefont
  {Lanco}},\ and\ \bibinfo {author} {\bibfnamefont {P.}~\bibnamefont
  {Senellart}},\ }\bibfield  {title} {\bibinfo {title} {Near-optimal
  single-photon sources in the solid state},\ }\href
  {https://doi.org/10.1038/nphoton.2016.23} {\bibfield  {journal} {\bibinfo
  {journal} {Nat. Photon.}\ }\textbf {\bibinfo {volume} {10}},\ \bibinfo
  {pages} {340} (\bibinfo {year} {2016})}\BibitemShut {NoStop}%
\bibitem [{\citenamefont {Meany}\ \emph {et~al.}(2015)\citenamefont {Meany},
  \citenamefont {Gr{\"a}fe}, \citenamefont {Heilmann}, \citenamefont
  {Perez-Leija}, \citenamefont {Gross}, \citenamefont {Steel}, \citenamefont
  {Withford},\ and\ \citenamefont {Szameit}}]{meany2015}%
  \BibitemOpen
  \bibfield  {author} {\bibinfo {author} {\bibfnamefont {T.}~\bibnamefont
  {Meany}}, \bibinfo {author} {\bibfnamefont {M.}~\bibnamefont {Gr{\"a}fe}},
  \bibinfo {author} {\bibfnamefont {R.}~\bibnamefont {Heilmann}}, \bibinfo
  {author} {\bibfnamefont {A.}~\bibnamefont {Perez-Leija}}, \bibinfo {author}
  {\bibfnamefont {S.}~\bibnamefont {Gross}}, \bibinfo {author} {\bibfnamefont
  {M.~J.}\ \bibnamefont {Steel}}, \bibinfo {author} {\bibfnamefont {M.~J.}\
  \bibnamefont {Withford}},\ and\ \bibinfo {author} {\bibfnamefont
  {A.}~\bibnamefont {Szameit}},\ }\bibfield  {title} {\bibinfo {title} {Laser
  written circuits for quantum photonics},\ }\href
  {https://doi.org/10.1002/lpor.201500061} {\bibfield  {journal} {\bibinfo
  {journal} {Laser Photon. Rev.}\ }\textbf {\bibinfo {volume} {9}},\ \bibinfo
  {pages} {363} (\bibinfo {year} {2015})}\BibitemShut {NoStop}%
\bibitem [{\citenamefont {Corrielli}\ \emph {et~al.}(2021)\citenamefont
  {Corrielli}, \citenamefont {Crespi},\ and\ \citenamefont
  {Osellame}}]{corrielli2021}%
  \BibitemOpen
  \bibfield  {author} {\bibinfo {author} {\bibfnamefont {G.}~\bibnamefont
  {Corrielli}}, \bibinfo {author} {\bibfnamefont {A.}~\bibnamefont {Crespi}},\
  and\ \bibinfo {author} {\bibfnamefont {R.}~\bibnamefont {Osellame}},\
  }\bibfield  {title} {\bibinfo {title} {Femtosecond laser micromachining for
  integrated quantum photonics},\ }\href
  {https://doi.org/10.1515/nanoph-2021-0419} {\bibfield  {journal} {\bibinfo
  {journal} {Nanophotonics}\ }\textbf {\bibinfo {volume} {10}},\ \bibinfo
  {pages} {3789} (\bibinfo {year} {2021})}\BibitemShut {NoStop}%
\bibitem [{\citenamefont {Rarity}\ \emph {et~al.}(1990)\citenamefont {Rarity},
  \citenamefont {Tapster}, \citenamefont {Jakeman}, \citenamefont {Larchuk},
  \citenamefont {Campos}, \citenamefont {Teich},\ and\ \citenamefont
  {Saleh}}]{rarity1990}%
  \BibitemOpen
  \bibfield  {author} {\bibinfo {author} {\bibfnamefont {J.}~\bibnamefont
  {Rarity}}, \bibinfo {author} {\bibfnamefont {P.}~\bibnamefont {Tapster}},
  \bibinfo {author} {\bibfnamefont {E.}~\bibnamefont {Jakeman}}, \bibinfo
  {author} {\bibfnamefont {T.}~\bibnamefont {Larchuk}}, \bibinfo {author}
  {\bibfnamefont {R.}~\bibnamefont {Campos}}, \bibinfo {author} {\bibfnamefont
  {M.}~\bibnamefont {Teich}},\ and\ \bibinfo {author} {\bibfnamefont
  {B.}~\bibnamefont {Saleh}},\ }\bibfield  {title} {\bibinfo {title}
  {Two-photon interference in a mach-zehnder interferometer},\ }\href
  {https://doi.org/10.1103/physrevlett.65.1348} {\bibfield  {journal} {\bibinfo
   {journal} {Phys. Rev. Lett.}\ }\textbf {\bibinfo {volume} {65}},\ \bibinfo
  {pages} {1348} (\bibinfo {year} {1990})}\BibitemShut {NoStop}%
\bibitem [{\citenamefont {Clements}\ \emph {et~al.}(2016)\citenamefont
  {Clements}, \citenamefont {Humphreys}, \citenamefont {Metcalf}, \citenamefont
  {Kolthammer},\ and\ \citenamefont {Walmsley}}]{clements2016}%
  \BibitemOpen
  \bibfield  {author} {\bibinfo {author} {\bibfnamefont {W.~R.}\ \bibnamefont
  {Clements}}, \bibinfo {author} {\bibfnamefont {P.~C.}\ \bibnamefont
  {Humphreys}}, \bibinfo {author} {\bibfnamefont {B.~J.}\ \bibnamefont
  {Metcalf}}, \bibinfo {author} {\bibfnamefont {W.~S.}\ \bibnamefont
  {Kolthammer}},\ and\ \bibinfo {author} {\bibfnamefont {I.~A.}\ \bibnamefont
  {Walmsley}},\ }\bibfield  {title} {\bibinfo {title} {Optimal design for
  universal multiport interferometers},\ }\href
  {https://doi.org/10.1364/optica.3.001460} {\bibfield  {journal} {\bibinfo
  {journal} {Optica}\ }\textbf {\bibinfo {volume} {3}},\ \bibinfo {pages}
  {1460} (\bibinfo {year} {2016})}\BibitemShut {NoStop}%
\bibitem [{\citenamefont {Laing}\ and\ \citenamefont
  {O'Brien}(2012)}]{laing2012}%
  \BibitemOpen
  \bibfield  {author} {\bibinfo {author} {\bibfnamefont {A.}~\bibnamefont
  {Laing}}\ and\ \bibinfo {author} {\bibfnamefont {J.~L.}\ \bibnamefont
  {O'Brien}},\ }\bibfield  {title} {\bibinfo {title} {Super-stable tomography
  of any linear optical device},\ }\href {https://arxiv.org/abs/1208.2868}
  {\bibfield  {journal} {\bibinfo  {journal} {arXiv preprint arXiv:1208.2868}\
  } (\bibinfo {year} {2012})}\BibitemShut {NoStop}%
\bibitem [{\citenamefont {Flamini}\ \emph {et~al.}(2015)\citenamefont
  {Flamini}, \citenamefont {Magrini}, \citenamefont {Rab}, \citenamefont
  {Spagnolo}, \citenamefont {D'Ambrosio}, \citenamefont {Mataloni},
  \citenamefont {Sciarrino}, \citenamefont {Zandrini}, \citenamefont {Crespi},
  \citenamefont {Ramponi} \emph {et~al.}}]{flamini2015}%
  \BibitemOpen
  \bibfield  {author} {\bibinfo {author} {\bibfnamefont {F.}~\bibnamefont
  {Flamini}}, \bibinfo {author} {\bibfnamefont {L.}~\bibnamefont {Magrini}},
  \bibinfo {author} {\bibfnamefont {A.~S.}\ \bibnamefont {Rab}}, \bibinfo
  {author} {\bibfnamefont {N.}~\bibnamefont {Spagnolo}}, \bibinfo {author}
  {\bibfnamefont {V.}~\bibnamefont {D'Ambrosio}}, \bibinfo {author}
  {\bibfnamefont {P.}~\bibnamefont {Mataloni}}, \bibinfo {author}
  {\bibfnamefont {F.}~\bibnamefont {Sciarrino}}, \bibinfo {author}
  {\bibfnamefont {T.}~\bibnamefont {Zandrini}}, \bibinfo {author}
  {\bibfnamefont {A.}~\bibnamefont {Crespi}}, \bibinfo {author} {\bibfnamefont
  {R.}~\bibnamefont {Ramponi}}, \emph {et~al.},\ }\bibfield  {title} {\bibinfo
  {title} {Thermally reconfigurable quantum photonic circuits at telecom
  wavelength by femtosecond laser micromachining},\ }\href
  {https://doi.org/10.1038/lsa.2015.127} {\bibfield  {journal} {\bibinfo
  {journal} {Light Sci. Appl.}\ }\textbf {\bibinfo {volume} {4}},\ \bibinfo
  {pages} {e354} (\bibinfo {year} {2015})}\BibitemShut {NoStop}%
\bibitem [{\citenamefont {Ceccarelli}\ \emph {et~al.}(2019)\citenamefont
  {Ceccarelli}, \citenamefont {Atzeni}, \citenamefont {Prencipe}, \citenamefont
  {Farinaro},\ and\ \citenamefont {Osellame}}]{ceccarelli2019}%
  \BibitemOpen
  \bibfield  {author} {\bibinfo {author} {\bibfnamefont {F.}~\bibnamefont
  {Ceccarelli}}, \bibinfo {author} {\bibfnamefont {S.}~\bibnamefont {Atzeni}},
  \bibinfo {author} {\bibfnamefont {A.}~\bibnamefont {Prencipe}}, \bibinfo
  {author} {\bibfnamefont {R.}~\bibnamefont {Farinaro}},\ and\ \bibinfo
  {author} {\bibfnamefont {R.}~\bibnamefont {Osellame}},\ }\bibfield  {title}
  {\bibinfo {title} {Thermal phase shifters for femtosecond laser written
  photonic integrated circuits},\ }\href
  {https://doi.org/10.1109/jlt.2019.2923126} {\bibfield  {journal} {\bibinfo
  {journal} {J. Light. Technol.}\ }\textbf {\bibinfo {volume} {37}},\ \bibinfo
  {pages} {4275} (\bibinfo {year} {2019})}\BibitemShut {NoStop}%
\bibitem [{\citenamefont {Arriola}\ \emph {et~al.}(2013)\citenamefont
  {Arriola}, \citenamefont {Gross}, \citenamefont {Jovanovic}, \citenamefont
  {Charles}, \citenamefont {Tuthill}, \citenamefont {Olaizola}, \citenamefont
  {Fuerbach},\ and\ \citenamefont {Withford}}]{arriola2013}%
  \BibitemOpen
  \bibfield  {author} {\bibinfo {author} {\bibfnamefont {A.}~\bibnamefont
  {Arriola}}, \bibinfo {author} {\bibfnamefont {S.}~\bibnamefont {Gross}},
  \bibinfo {author} {\bibfnamefont {N.}~\bibnamefont {Jovanovic}}, \bibinfo
  {author} {\bibfnamefont {N.}~\bibnamefont {Charles}}, \bibinfo {author}
  {\bibfnamefont {P.~G.}\ \bibnamefont {Tuthill}}, \bibinfo {author}
  {\bibfnamefont {S.~M.}\ \bibnamefont {Olaizola}}, \bibinfo {author}
  {\bibfnamefont {A.}~\bibnamefont {Fuerbach}},\ and\ \bibinfo {author}
  {\bibfnamefont {M.~J.}\ \bibnamefont {Withford}},\ }\bibfield  {title}
  {\bibinfo {title} {Low bend loss waveguides enable compact, efficient 3d
  photonic chips},\ }\href {https://doi.org/10.1364/oe.21.002978} {\bibfield
  {journal} {\bibinfo  {journal} {Opt. Express}\ }\textbf {\bibinfo {volume}
  {21}},\ \bibinfo {pages} {2978} (\bibinfo {year} {2013})}\BibitemShut
  {NoStop}%
\bibitem [{\citenamefont {Nowak}\ \emph {et~al.}(2014)\citenamefont {Nowak},
  \citenamefont {Portalupi}, \citenamefont {Giesz}, \citenamefont {Gazzano},
  \citenamefont {{Dal Savio}}, \citenamefont {Braun}, \citenamefont {Karrai},
  \citenamefont {Arnold}, \citenamefont {Lanco}, \citenamefont {Sagnes},
  \citenamefont {Lema{\^{i}}tre},\ and\ \citenamefont
  {Senellart}}]{Nowak2014a}%
  \BibitemOpen
  \bibfield  {author} {\bibinfo {author} {\bibfnamefont {A.~K.}\ \bibnamefont
  {Nowak}}, \bibinfo {author} {\bibfnamefont {S.~L.}\ \bibnamefont
  {Portalupi}}, \bibinfo {author} {\bibfnamefont {V.}~\bibnamefont {Giesz}},
  \bibinfo {author} {\bibfnamefont {O.}~\bibnamefont {Gazzano}}, \bibinfo
  {author} {\bibfnamefont {C.}~\bibnamefont {{Dal Savio}}}, \bibinfo {author}
  {\bibfnamefont {P.~F.}\ \bibnamefont {Braun}}, \bibinfo {author}
  {\bibfnamefont {K.}~\bibnamefont {Karrai}}, \bibinfo {author} {\bibfnamefont
  {C.}~\bibnamefont {Arnold}}, \bibinfo {author} {\bibfnamefont
  {L.}~\bibnamefont {Lanco}}, \bibinfo {author} {\bibfnamefont
  {I.}~\bibnamefont {Sagnes}}, \bibinfo {author} {\bibfnamefont
  {A.}~\bibnamefont {Lema{\^{i}}tre}},\ and\ \bibinfo {author} {\bibfnamefont
  {P.}~\bibnamefont {Senellart}},\ }\bibfield  {title} {\bibinfo {title}
  {{Deterministic and electrically tunable bright single-photon source}},\
  }\href {https://doi.org/10.1038/ncomms4240} {\bibfield  {journal} {\bibinfo
  {journal} {Nat. Commun.}\ }\textbf {\bibinfo {volume} {5}},\ \bibinfo {pages}
  {1} (\bibinfo {year} {2014})}\BibitemShut {NoStop}%
\bibitem [{\citenamefont {Loredo}\ \emph {et~al.}(2016)\citenamefont {Loredo},
  \citenamefont {Zakaria}, \citenamefont {Somaschi}, \citenamefont {Anton},
  \citenamefont {de~Santis}, \citenamefont {Giesz}, \citenamefont {Grange},
  \citenamefont {Broome}, \citenamefont {Gazzano}, \citenamefont {Coppola},
  \citenamefont {Sagnes}, \citenamefont {Lemaitre}, \citenamefont {Auffeves},
  \citenamefont {Senellart}, \citenamefont {Almeida},\ and\ \citenamefont
  {White}}]{Loredo2016}%
  \BibitemOpen
  \bibfield  {author} {\bibinfo {author} {\bibfnamefont {J.~C.}\ \bibnamefont
  {Loredo}}, \bibinfo {author} {\bibfnamefont {N.~A.}\ \bibnamefont {Zakaria}},
  \bibinfo {author} {\bibfnamefont {N.}~\bibnamefont {Somaschi}}, \bibinfo
  {author} {\bibfnamefont {C.}~\bibnamefont {Anton}}, \bibinfo {author}
  {\bibfnamefont {L.}~\bibnamefont {de~Santis}}, \bibinfo {author}
  {\bibfnamefont {V.}~\bibnamefont {Giesz}}, \bibinfo {author} {\bibfnamefont
  {T.}~\bibnamefont {Grange}}, \bibinfo {author} {\bibfnamefont {M.~A.}\
  \bibnamefont {Broome}}, \bibinfo {author} {\bibfnamefont {O.}~\bibnamefont
  {Gazzano}}, \bibinfo {author} {\bibfnamefont {G.}~\bibnamefont {Coppola}},
  \bibinfo {author} {\bibfnamefont {I.}~\bibnamefont {Sagnes}}, \bibinfo
  {author} {\bibfnamefont {A.}~\bibnamefont {Lemaitre}}, \bibinfo {author}
  {\bibfnamefont {A.}~\bibnamefont {Auffeves}}, \bibinfo {author}
  {\bibfnamefont {P.}~\bibnamefont {Senellart}}, \bibinfo {author}
  {\bibfnamefont {M.~P.}\ \bibnamefont {Almeida}},\ and\ \bibinfo {author}
  {\bibfnamefont {A.~G.}\ \bibnamefont {White}},\ }\bibfield  {title} {\bibinfo
  {title} {Scalable performance in solid-state single-photon sources},\ }\href
  {https://doi.org/10.1364/OPTICA.3.000433} {\bibfield  {journal} {\bibinfo
  {journal} {Optica}\ }\textbf {\bibinfo {volume} {3}},\ \bibinfo {pages} {433}
  (\bibinfo {year} {2016})}\BibitemShut {NoStop}%
\bibitem [{\citenamefont {Spagnolo}\ \emph {et~al.}(2013)\citenamefont
  {Spagnolo}, \citenamefont {Vitelli}, \citenamefont {Sansoni}, \citenamefont
  {Maiorino}, \citenamefont {Mataloni}, \citenamefont {Sciarrino},
  \citenamefont {Brod}, \citenamefont {Galvao}, \citenamefont {Crespi},
  \citenamefont {Ramponi} \emph {et~al.}}]{spagnolo2013}%
  \BibitemOpen
  \bibfield  {author} {\bibinfo {author} {\bibfnamefont {N.}~\bibnamefont
  {Spagnolo}}, \bibinfo {author} {\bibfnamefont {C.}~\bibnamefont {Vitelli}},
  \bibinfo {author} {\bibfnamefont {L.}~\bibnamefont {Sansoni}}, \bibinfo
  {author} {\bibfnamefont {E.}~\bibnamefont {Maiorino}}, \bibinfo {author}
  {\bibfnamefont {P.}~\bibnamefont {Mataloni}}, \bibinfo {author}
  {\bibfnamefont {F.}~\bibnamefont {Sciarrino}}, \bibinfo {author}
  {\bibfnamefont {D.~J.}\ \bibnamefont {Brod}}, \bibinfo {author}
  {\bibfnamefont {E.~F.}\ \bibnamefont {Galvao}}, \bibinfo {author}
  {\bibfnamefont {A.}~\bibnamefont {Crespi}}, \bibinfo {author} {\bibfnamefont
  {R.}~\bibnamefont {Ramponi}}, \emph {et~al.},\ }\bibfield  {title} {\bibinfo
  {title} {General rules for bosonic bunching in multimode interferometers},\
  }\href {https://doi.org/10.1103/physrevlett.111.130503} {\bibfield  {journal}
  {\bibinfo  {journal} {Phys. Rev. Lett.}\ }\textbf {\bibinfo {volume} {111}},\
  \bibinfo {pages} {130503} (\bibinfo {year} {2013})}\BibitemShut {NoStop}%
\bibitem [{\citenamefont {Dousse}\ \emph {et~al.}(2009)\citenamefont {Dousse},
  \citenamefont {Suffczy{\'n}ski}, \citenamefont {Braive}, \citenamefont
  {Miard}, \citenamefont {Lema{\^\i}tre}, \citenamefont {Sagnes}, \citenamefont
  {Lanco}, \citenamefont {Bloch}, \citenamefont {Voisin},\ and\ \citenamefont
  {Senellart}}]{dousse2009}%
  \BibitemOpen
  \bibfield  {author} {\bibinfo {author} {\bibfnamefont {A.}~\bibnamefont
  {Dousse}}, \bibinfo {author} {\bibfnamefont {J.}~\bibnamefont
  {Suffczy{\'n}ski}}, \bibinfo {author} {\bibfnamefont {R.}~\bibnamefont
  {Braive}}, \bibinfo {author} {\bibfnamefont {A.}~\bibnamefont {Miard}},
  \bibinfo {author} {\bibfnamefont {A.}~\bibnamefont {Lema{\^\i}tre}}, \bibinfo
  {author} {\bibfnamefont {I.}~\bibnamefont {Sagnes}}, \bibinfo {author}
  {\bibfnamefont {L.}~\bibnamefont {Lanco}}, \bibinfo {author} {\bibfnamefont
  {J.}~\bibnamefont {Bloch}}, \bibinfo {author} {\bibfnamefont
  {P.}~\bibnamefont {Voisin}},\ and\ \bibinfo {author} {\bibfnamefont
  {P.}~\bibnamefont {Senellart}},\ }\bibfield  {title} {\bibinfo {title}
  {Scalable implementation of strongly coupled cavity-quantum dot devices},\
  }\href {https://doi.org/10.1063/1.3100781} {\bibfield  {journal} {\bibinfo
  {journal} {Appl. Phys. Lett.}\ }\textbf {\bibinfo {volume} {94}},\ \bibinfo
  {pages} {121102} (\bibinfo {year} {2009})}\BibitemShut {NoStop}%
\bibitem [{\citenamefont {Galv\~ao}\ and\ \citenamefont
  {Brod}(2020)}]{brod2020}%
  \BibitemOpen
  \bibfield  {author} {\bibinfo {author} {\bibfnamefont {E.~F.}\ \bibnamefont
  {Galv\~ao}}\ and\ \bibinfo {author} {\bibfnamefont {D.~J.}\ \bibnamefont
  {Brod}},\ }\bibfield  {title} {\bibinfo {title} {Quantum and classical bounds
  for two-state overlaps},\ }\href
  {https://doi.org/10.1103/PhysRevA.101.062110} {\bibfield  {journal} {\bibinfo
   {journal} {Phys. Rev. A}\ }\textbf {\bibinfo {volume} {101}},\ \bibinfo
  {pages} {062110} (\bibinfo {year} {2020})}\BibitemShut {NoStop}%
\bibitem [{\citenamefont {Oszmaniec}\ and\ \citenamefont
  {Brod}(2018)}]{Oszm18}%
  \BibitemOpen
  \bibfield  {author} {\bibinfo {author} {\bibfnamefont {M.}~\bibnamefont
  {Oszmaniec}}\ and\ \bibinfo {author} {\bibfnamefont {D.~J.}\ \bibnamefont
  {Brod}},\ }\bibfield  {title} {\bibinfo {title} {Classical simulation of
  photonic linear optics with lost particles},\ }\href
  {https://doi.org/10.1088/1367-2630/aadfa8} {\bibfield  {journal} {\bibinfo
  {journal} {New J. Phys.}\ }\textbf {\bibinfo {volume} {20}},\ \bibinfo
  {pages} {092002} (\bibinfo {year} {2018})}\BibitemShut {NoStop}%
\bibitem [{\citenamefont {Tichy}(2015)}]{tichy2015}%
  \BibitemOpen
  \bibfield  {author} {\bibinfo {author} {\bibfnamefont {M.~C.}\ \bibnamefont
  {Tichy}},\ }\bibfield  {title} {\bibinfo {title} {Sampling of partially
  distinguishable bosons and the relation to the multidimensional permanent},\
  }\href {https://doi.org/10.1103/physreva.91.022316} {\bibfield  {journal}
  {\bibinfo  {journal} {Phys. Rev. A}\ }\textbf {\bibinfo {volume} {91}},\
  \bibinfo {pages} {022316} (\bibinfo {year} {2015})}\BibitemShut {NoStop}%
\bibitem [{\citenamefont {Renema}\ \emph {et~al.}(2018)\citenamefont {Renema},
  \citenamefont {Menssen}, \citenamefont {Clements}, \citenamefont {Triginer},
  \citenamefont {Kolthammer},\ and\ \citenamefont {Walmsley}}]{Rene18}%
  \BibitemOpen
  \bibfield  {author} {\bibinfo {author} {\bibfnamefont {J.~J.}\ \bibnamefont
  {Renema}}, \bibinfo {author} {\bibfnamefont {A.}~\bibnamefont {Menssen}},
  \bibinfo {author} {\bibfnamefont {W.~R.}\ \bibnamefont {Clements}}, \bibinfo
  {author} {\bibfnamefont {G.}~\bibnamefont {Triginer}}, \bibinfo {author}
  {\bibfnamefont {W.~S.}\ \bibnamefont {Kolthammer}},\ and\ \bibinfo {author}
  {\bibfnamefont {I.~A.}\ \bibnamefont {Walmsley}},\ }\bibfield  {title}
  {\bibinfo {title} {Efficient classical algorithm for boson sampling with
  partially distinguishable photons},\ }\href
  {https://doi.org/10.1103/PhysRevLett.120.220502} {\bibfield  {journal}
  {\bibinfo  {journal} {Phys. Rev. Lett.}\ }\textbf {\bibinfo {volume} {120}},\
  \bibinfo {pages} {220502} (\bibinfo {year} {2018})}\BibitemShut {NoStop}%
\bibitem [{\citenamefont {Moylett}\ \emph {et~al.}(2020)\citenamefont
  {Moylett}, \citenamefont {Garcia-Patron}, \citenamefont {Renema},\ and\
  \citenamefont {Turner}}]{Moyl19}%
  \BibitemOpen
  \bibfield  {author} {\bibinfo {author} {\bibfnamefont {A.}~\bibnamefont
  {Moylett}}, \bibinfo {author} {\bibfnamefont {R.}~\bibnamefont
  {Garcia-Patron}}, \bibinfo {author} {\bibfnamefont {J.~J.}\ \bibnamefont
  {Renema}},\ and\ \bibinfo {author} {\bibfnamefont {P.~S.}\ \bibnamefont
  {Turner}},\ }\bibfield  {title} {\bibinfo {title} {Classically simulating
  near-term partially-distinguishable and lossy boson sampling},\ }\href
  {https://doi.org/10.1088/2058-9565/ab5555} {\bibfield  {journal} {\bibinfo
  {journal} {Quantum Sci. Technol.}\ }\textbf {\bibinfo {volume} {5}},\
  \bibinfo {pages} {010501} (\bibinfo {year} {2020})}\BibitemShut {NoStop}%
\end{thebibliography}

%


\clearpage

\appendix
	
\begin{center}
\bf APPENDIX
\end{center}

\section{Influence of the phase terms}
\label{sec:phases}

Physical phase shifts, placed in one of the internal optical paths of a multimode interferometer, can affect the measurable output state in a way that is not straightforward. In addition, the observation that distinct phase shifters affect in a measurable way the output state, does not mean that they are acting on independent parameters of the unitary transformation of the multimode device.

\begin{figure}
\centering
\includegraphics[scale=1]{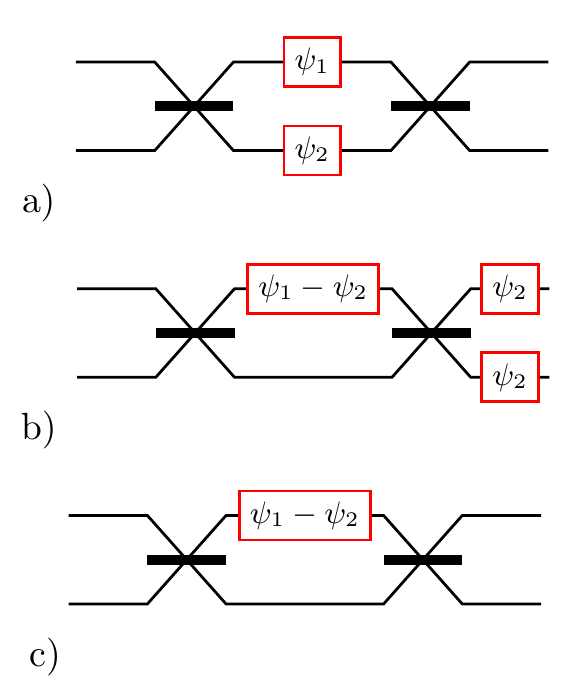}
\caption{\label{fig:MZI} The Mach-Zehnder interferometers (a) and (b) are described by precisely the same unitary matrix $U_\text{MZI}$. The interferometer in (c) is described instead by an equivalent matrix $U'_\text{MZI} = C \cdot U_\text{MZI}$ where $C$ is a diagonal matrix with only phase terms.}
\end{figure}

Here we exemplify these considerations in the simple case of a two-arm Mach-Zehnder interferometer, as the one depicted in Fig.~\ref{fig:MZI}a. The device is composed of two symmetric beam-splitters and two different phase plates in the internal arms. If we model the beam-splitters with the matrix:
\begin{equation}
U_\text{BS} = \frac{\sqrt{2}}{2} \begin{bmatrix} 1 & i \\ i & 1 \end{bmatrix}
\label{eq:Ubs}
\end{equation}
the matrix of the full interferometer is calculated as:
\begin{align}
U_\text{MZI} &= U_\text{BS} \cdot \begin{bmatrix} e^{i \psi_1} & 0 \\ 0 & e^{i \psi_2} \end{bmatrix} \cdot U_\text{BS} =\notag\\
&= \frac 1 2 \begin{bmatrix} e^{i \psi_1} - e^{i \psi_2} & i\left( e^{i \psi_1} +  e^{i \psi_2}\right) \\ i\left(e^{i \psi_1} + i e^{i \psi_2}\right) &  - e^{i \psi_1} + e^{i \psi_2} \end{bmatrix}
\end{align}
It is clear that, in general, both varying $\psi_1$ or $\psi_2$ will have some measurable influence on the output state of the interferometer; in particular, a change in either of the two phases will produce interference fringes.

Let us now calculate the matrix of the other interferometer depicted in panel (b). Actually, one would observe that is exactly the same as $U_\text{MZI}$:
\begin{align}
&\begin{bmatrix} e^{i \psi_2} & 0 \\ 0 & e^{i \psi_2} \end{bmatrix} \cdot U_\text{BS} \cdot \begin{bmatrix} e^{i (\psi_1-\psi_2)} & 0 \\ 0 & 1 \end{bmatrix} \cdot U_\text{BS} =\notag\\
&=\frac 1 2 \begin{bmatrix} e^{i \psi_1} - e^{i \psi_2} & i\left( e^{i \psi_1} +  e^{i \psi_2}\right) \\ i\left(e^{i \psi_1} + i e^{i \psi_2}\right) &  - e^{i \psi_1} + e^{i \psi_2} \end{bmatrix}=U_\text{MZI}
\end{align}

The perfect correspondence between the two configurations of Fig.~\ref{fig:MZI}a and b can be also understood by analyzing the phase delays accumulated along the optical paths. 
All possible paths result in either $\psi_1$ or $\psi_2$ delay in both configurations.

At this point, one would comment that the phases $\psi_2$ at the outputs of the second Mach-Zehnder may be removed without altering the effective circuit operation on Fock states. This means that the interferometers in Fig.~\ref{fig:MZI}a-b are both equivalent to the one represented in Fig.~\ref{fig:MZI}c, which is described by the matrix:
\begin{equation}
U'_\text{MZI} = \frac{1}{2} \begin{bmatrix} e^{i (\psi_1-\psi_2)} - 1 & i\left( e^{i (\psi_1-\psi_2)} + 1\right) \\ i\left( e^{i (\psi_1-\psi_2)} + 1\right) &  -e^{i (\psi_1-\psi_2)} + 1 \end{bmatrix}
\end{equation}
This also means that what is actually affecting the operation of this interferometer is the \emph{phase difference} $\psi_1 - \psi_2$ and not the two individual phases. In other terms, there is only one degree of freedom available for both of these elements. 

\section{Equivalent cyclic interferometers}
\label{sec:simplification}

We discuss in this Appendix how to transform the general layout of C.I. shown in Fig.~\ref{fig:interf1}b into the simplified one of Fig.~\ref{fig:interf1}c.

\begin{figure*}
\centering
\includegraphics[scale=1]{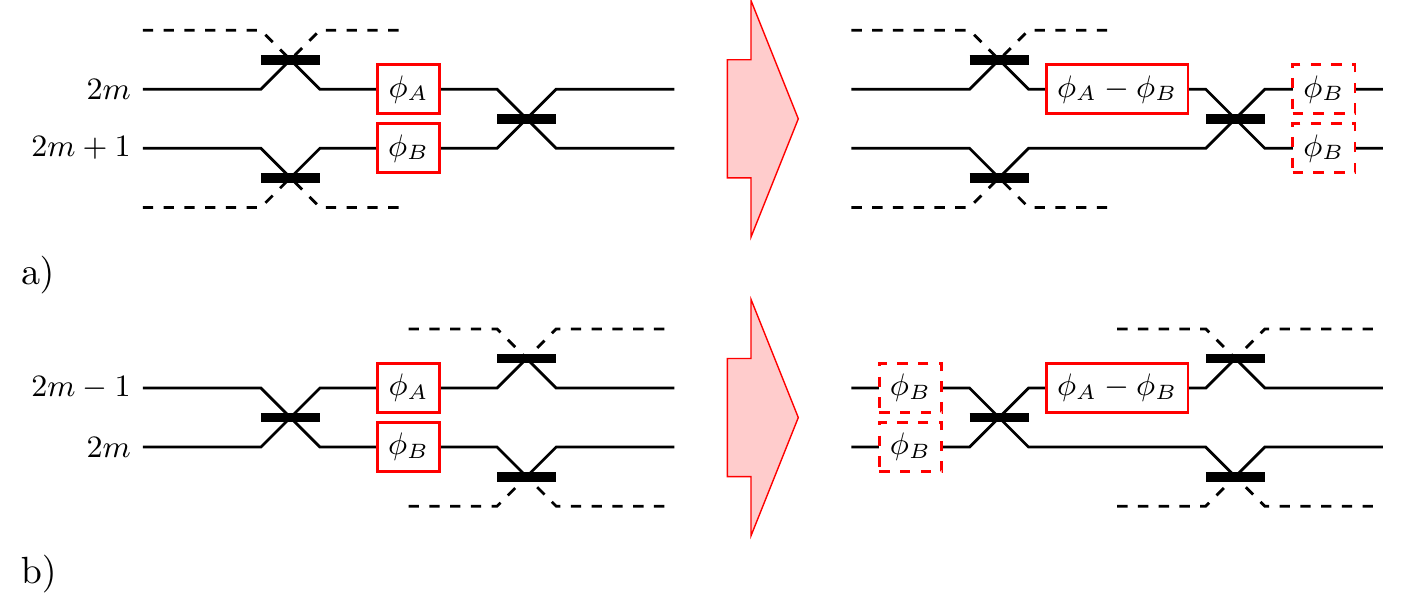}
\caption{\label{fig:simplification} Procedures for reducing the number of phase shifters in the device of Fig.~\ref{fig:interf1}a, without changing its output distribution when Fock states are injected at the inputs. The procedures allow to set to zero the phase term placed on one of the internal arms, by modifying the phase on the upper adjacent arm and by adding phase shifts at the inputs or at the outputs. (a) and (b) hold respectively for odd and even arms.}
\end{figure*}

We consider the two procedures schematized in Fig.~\ref{fig:simplification}. In panel (a) it is shown how the phase delay $\phi_B$, placed on an odd mode $2m+1$, can be removed (i.e. the phase on that arm can be set to zero), without any change in the matrix of the device. To do so, the phase $\phi_A$ present on mode $2m$ is changed to $\phi_A-\phi_B$, and two phase shifts equal to $\phi_B$ are added at the outputs $2m$ and $2m+1$.
One observes that, for any chosen couple of input and output ports, a photon travelling in the circuit undergoes the same overall phase delay in the two configurations. Since beam splitter operations have not changed, this means that the two configurations are described by precisely the same unitary matrix. 
The procedure shown in Fig.~\ref{fig:simplification}b is analogous to the one just described, but applies to a phase delay $\phi_B$ placed on an even mode $2m$. In this case, the phase shifter $\phi_A$ already present on mode $2m-1$ is changed to $\phi_A-\phi_B$, and two phase shifters $\phi_B$ are added at the inputs $2m-1$ and $2m$.

These two procedures can be applied alternately starting from the $N$-th mode up to the second mode of the C.I. in Fig.~\ref{fig:interf1}b, producing an interferometer with all the phase terms on the internal arms set to zero, except for the phase $\alpha_1$ on the first arm.
Given the iterative simplification procedures that has been operated, the phase term $\alpha_1$ contains an algebraic sum of all the phases $\phi_i$ initially considered in the interferometer, as in Eq.~\eqref{eq:alpha}.
Then, the phase shifters at the inputs and at the outputs, represented with dashed contours, can be removed without changing the output distribution of the device (see also the example in Appendix~\ref{sec:phases}). This proves that the interferometer in Fig.~\ref{fig:interf1}c is equivalent to the one in Fig.~\ref{fig:interf1}b, in the sense of Definition~\ref{def:equivalent}.

\section{Odd-modes to odd-modes\\ detection probability}
\label{sec:permanents}

We consider in this Appendix an experiment conducted using a C.I. with $N=2n$ modes, in which $n$ photons are injected one per each odd input mode, and $n$-photon coincidence detection is operated at the output, detecting one photon per each odd output mode.

\subsection{Indistinguishable photons}

If the $n$-photons are perfectly identical, the probability to detect this output state is given by \cite{spagnolo2013}:
\begin{equation} \label{eq:permanent}
P_{\vec{g},\vec{h}} = \frac{|\perm S_{\vec{g},\vec{h}}|^2}{\mu_1! \mu_2! \ldots \mu_N! \nu_1! \nu_2! \ldots 	\nu_N!}
\end{equation}
where $\mu_i$ and $\nu_i$ are the number of particles present in mode $i$ in the $g$ and $h$ states respectively. $S_{\vec{g},\vec{h}}$ is the scattering matrix with elements $S_{i,j} = U_{h_i,g_j}$ (where $U$ is the unitary matrix of the interferometer) and $\perm S$ denotes the permanent of the matrix $S$.
 
The elements of the $n \times n$ scattering matrix $S$ can be worked out by inspection of the interferometer layout, considering the attenuation and phase delays undergone by a photon in all the relevant optical paths. In detail, for the considered input and output states, the element $S_{r,c}$ (with $1\leq r \leq n$ and $1\leq c \leq n$) is related to the path from the input port $2c-1$ to the output port $2r-1$ of the full device. Among the different equivalent interferometers, we refer to the one represented in Fig.~\ref{fig:interf1}c. 
 
A photon injected in the $(2m-1)$-th mode can reach the $(2m+1)$-th mode undergoing two 50\% transmissions on the first and second beam splitters, each giving a phase delay equal to $i = e^{i \pi/2}$ (we describe the beam splitters as in Eq.~\eqref{eq:Ubs}). Thus:
\begin{equation}
S_{m+1,m} = \frac{1}{2} i^2 = -\frac{1}{2}
\quad m \neq n \quad  m \neq 1
\end{equation}
Alternatively, a photon from the same input may exit from the $(2m-1)$-th mode undergoing two 50\% reflections (each giving a null phase delay, according to our beam splitter model):
\begin{equation}
S_{m,m} = \frac{1}{2},\quad  m \neq 1
\end{equation}
All other elements of the scattering matrix associated to this input are vanishing, because there are no other odd modes connected to it.

To complete the picture, we take into account the special case given by the first and by the $(N-1)$-th inputs of the interferometer, which produce the following elements that do not obey to the previous rules:
\begin{align}
S_{1,1} &= \frac{1}{2} e^{i \alpha_1} & S_{1,n} &= -\frac{1}{2}
\end{align}
Finally, the scattering matrix $S$ is given by:
\begin{equation}
S = \frac 1 2 \underbrace{\begin{bmatrix}
e^{i \alpha_1} & & & & & -1\\
        -1 & 1 & & & &\\
           & -1 & 1 & &\\
           & &\ddots & \ddots & &\\
           &  & &\ddots & \ddots &\\
           & & & & -1 & 1
\end{bmatrix}}_{n \times n}
\end{equation}
which is essentially a bidiagonal matrix with an additional $S_{1,n}$ element.
The permanent of $S$ can be developed as follows, according to the Laplace formula (adapted to permanents):
\begin{align}
&\perm S = \frac{e^{i \alpha_1}}{2}  \cdot \perm \frac{1}{2} \underbrace{\begin{bmatrix}
 1 & & & &\\
-1 & 1 & &\\
   &\ddots & \ddots & &\\
   & &\ddots & \ddots &\\
   & & & -1 & 1
\end{bmatrix}}_{(n-1)\times (n-1)} +\notag\\
&-\frac{1}{2}\cdot \perm \frac{1}{2} \underbrace{\begin{bmatrix}
 -1 & 1 & & &\\
 & -1 & 1 &\\
   & & \ddots & \ddots & \\
   & & & \ddots & 1\\
   & & & & -1
\end{bmatrix}}_{(n-1)\times (n-1)}
\end{align}
Now, by applying recursively the Laplace formula it is not difficult to show that:
\begin{equation}
\perm \frac{1}{2} \begin{bmatrix}
 1 & & & &\\
-1 & 1 & &\\
   &\ddots & \ddots & &\\
   & &\ddots & \ddots &\\
   & & & -1 & 1
\end{bmatrix} = \left(\frac{1}{2}\right)^{n-1}
\end{equation}
and
\begin{equation}
\perm \frac{1}{2}\begin{bmatrix}
 -1 & 1 & & &\\
 & -1 & 1 &\\
   & & \ddots & \ddots & \\
   & & & \ddots & 1\\
   & & & & -1
\end{bmatrix} = \left(-\frac{1}{2}\right)^{n-1}
\end{equation}
where $n-1$ is the dimension of the matrix.
Hence, we obtain:
\begin{equation}
\perm S = \frac{1}{2^n} \cdot \left( e^{i \alpha_1} + (-1)^n \right)
\end{equation}
As a consequence, the probability to detect the output state $\vec h = (1,3,5,\dots)$, given the input state $\vec g = (1,3,5,\dots)$, has the expression given by Eq.~\eqref{eq:Pind} and Proposition~\ref{prop:prob135135} is proven.

\subsection{Distinguishable photons}

If the photons injected in the odd modes are completely distinguishable, the probability to detect them at the output again on the odd modes is given by \cite{spagnolo2013}:
\begin{equation}
P_\mathrm{dist} = \perm \left\lbrace|S_{i,j}|^2\right\rbrace 
\end{equation}

Thus, we need to calculate:
\begin{equation}
\perm \left\lbrace|S_{i,j}|^2\right\rbrace = \perm \frac{1}{2^2} \underbrace{\begin{bmatrix}
1 & & & & & 1\\
        1 & 1 & & & &\\
           & 1 & 1 & &\\
           & &\ddots & \ddots & &\\
           &  & &\ddots & \ddots &\\
           & & & & 1 & 1
\end{bmatrix}}_{n \times n} 
\end{equation}
Proceeding similarly to the previous case we have:
\begin{align}
&\perm \left\lbrace|S_{i,j}|^2\right\rbrace = \frac{1}{4}  \cdot \perm \frac{1}{4} \underbrace{\begin{bmatrix}
 1 & & & &\\
1 & 1 & &\\
   &\ddots & \ddots & &\\
   & &\ddots & \ddots &\\
   & & & 1 & 1
\end{bmatrix}}_{(n-1)\times (n-1)} +\notag\\
&+\frac{1}{4}\cdot \perm \frac{1}{4} \underbrace{\begin{bmatrix}
 1 & 1 & & &\\
 & 1 & 1 &\\
   & & \ddots & \ddots & \\
   & & & \ddots & 1\\
   & & & & 1
\end{bmatrix}}_{(n-1)\times (n-1)} \notag\\
\end{align}
where
\begin{align}
&\perm \frac{1}{4} \begin{bmatrix}
 1 & & & &\\
1 & 1 & &\\
   &\ddots & \ddots & &\\
   & &\ddots & \ddots &\\
   & & & 1 & 1
\end{bmatrix} = \notag\\
&= \perm \frac{1}{4}\begin{bmatrix}
 1 & 1 & & &\\
 & 1 & 1 &\\
   & & \ddots & \ddots & \\
   & & & \ddots & 1\\
   & & & & 1
\end{bmatrix} = \left(\frac{1}{4}\right)^{n-1}
\end{align}
Therefore, we obtain:
\begin{equation}
\perm \left\lbrace|S_{i,j}|^2\right\rbrace = 2 \cdot \frac{1}{4} \cdot  \left(\frac{1}{4}\right)^{n-1} = \frac{1}{2^{2n-1}}
\end{equation}
as in Eq.~\eqref{eq:pOddDist}.

\section{Generic {\it n}-photon interference fringes}
\label{sec:states}

We prove here Proposition~\ref{prop:allStates}, which we report again below for convenience:

\textit{
Given a $N=2n$ mode C.I., an $n$-photon interference fringe as a function of the internal phases can be measured \emph{if and only if} the input state has precisely one photon per each pair of input ports (1,2)-{\ldots}-($2m-1$,$2m$)-{\ldots}, and the output state has precisely one photon per each pair of output ports (2,3)-{\ldots}-($2m$,$2m+1$)-{\ldots}-(1,$N$).  In particular, for each input-output combination satisfying this rule, the detection probability takes the form:
\begin{equation}
P = \frac{1}{2^{2n-1}} \left(1 + (-1)^{n+p+q}\cdot \cos \alpha_1\right)
\end{equation}
where $p$ is the number of occupied even modes in the input state, and $q$ is the number of occupied even modes in the output state.
}

First, we prove that an input state with $n$ photons that does not satisfy the above described condition will not produce any interference fringe, as a function of the internal phases of the C.I..

Let us indeed consider an input state for which at least a given pair of inputs ($2m-1$,$2m$) is empty. Then we can choose, as a model for the interferometer, the one having a single phase shifter $\alpha_{2m-1}$ on the $(2m-1)$-th mode: no photon of the input state would propagate across this arm, which means that varying its phase value would not give any interference fringe. 
An analogous reasoning can be made, considering an output state with at least a given pair of outputs ($2m$,$2m+1$) empty. Choosing indeed the model of the interferometer that has a single phase shifter on mode $2m+1$, we observe that this phase shifter would not be able to influence any of the photons present in the output state. 

Second, we prove that all the mentioned input and output states provide multi-photon interference fringes.

We know already (see Proposition~\ref{prop:prob135135}, proved in Appendix~\ref{sec:permanents}) that the couple of states $\vec g = \vec h =  (1,3,5,\dots)$ produce interference fringes as a function of the internal phases.
Let us consider what happens when we change the position of one input photon of $\vec g$ from an odd mode to the following even mode (i.e., from mode $2m-1$ to mode $2m$).  This pair of odd and even input modes of the interferometer correspond respectively to the upper and lower input modes of the same beam splitter, among the ones composing the first layer. 

\begin{figure}[p!]
\centering
\includegraphics[scale=1]{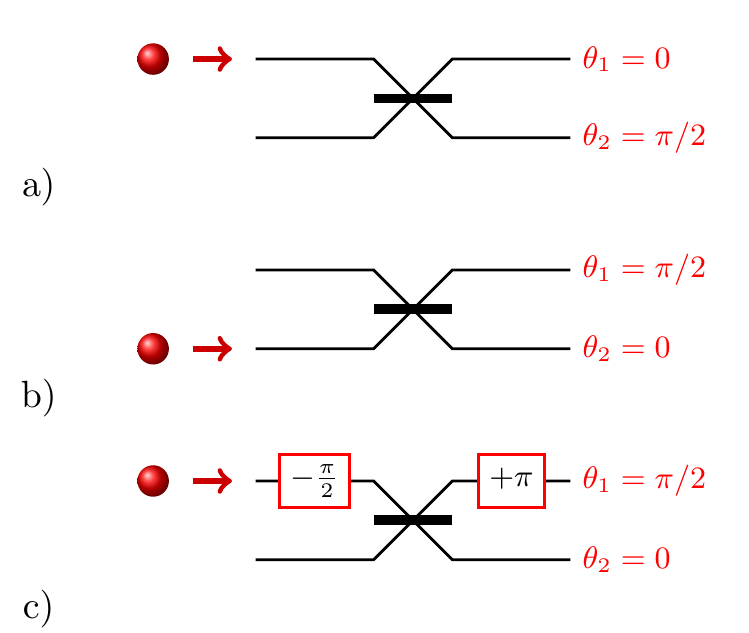}
\caption{\label{fig:exchangeInput} (a) A photon that enters the upper port of a beam splitter, according to our  modelling (Eq.~\eqref{eq:Ubs}), acquires $\theta_1=0$ phase delay when exiting on the upper port, and $\theta_1=\pi/2$ phase delay when exiting on the lower port. (b) A photon that enters the lower port has the symmetric behaviour. (c) Suitable phase shifters, placed on the input and output ports, make a photon entering the upper port to acquire the same phase delays as if it entered the lower one.}
\end{figure}

Since the beam splitter is balanced, the probability for a photon to exit in the upper or lower port of the beam splitter is the same, independently of the port from which it has entered. However, depending on the input, the phase acquired is different, according to the phase terms of the elements of the unitary matrix~\eqref{eq:FockState}, as shown in Fig.~\ref{fig:exchangeInput}a-b. If we place suitable phase shifters as in Fig.~\ref{fig:exchangeInput}c, a photon entering the upper (odd) mode acquires at the two output arms the same phase delays as if it entered the lower (even) mode. One of such phase shifters, having value $-\pi/2$, is placed directly at the input port and cannot affect in any way the probability distribution of the photons at output of the complete interferometer. The other phase shifter, having value $\pi$, is instead added in the internal paths of the interferometer. In practice, exchanging the position of one of the input photons (within the same pair of allowed inputs) is equivalent to adding $\pi$ to the internal phase $\alpha_1$ in Eq.~\eqref{eq:Pind}; we further note that adding $\pi$ to $\alpha_1$ is equivalent to a change of the sign of $\cos \alpha_1$ in the formula.

\begin{figure}[p!]
\centering
\includegraphics[scale=1]{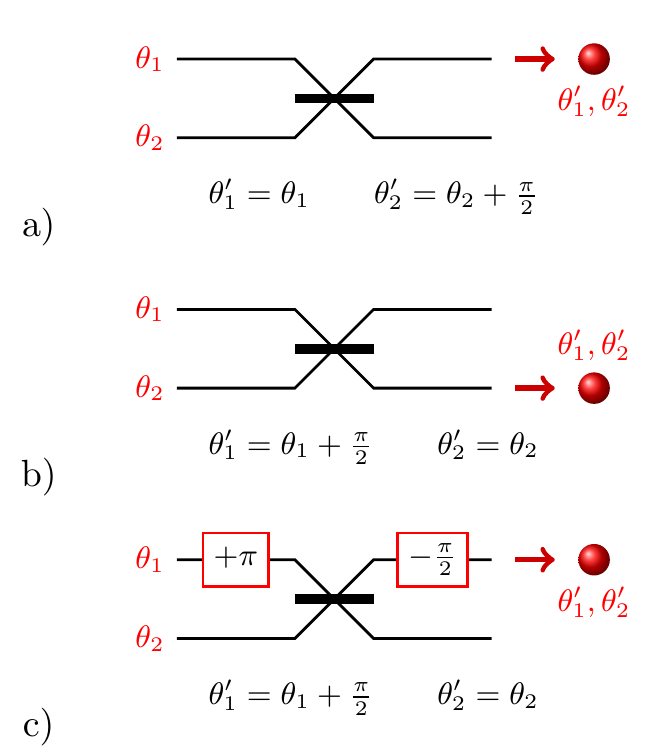}
\caption{\label{fig:exchangeOutput} (a) A photon detected at the upper output port of a beam splitter has acquired a total phase $\theta'_1 = \theta_1$ if it had entered the beam splitter from the upper input port carrying a phase $\theta_1$, or a phase $\theta'_1 = \theta_1$ if it had entered the beam splitter from the lower input port carrying a phase $\theta_2$. (b) Symmetric considerations can be made for a photon detected at the lower output port. (c) Suitable phase shifters can be added, to make the phases acquired by a photon detected on the upper output port equal to the ones that would have been acquired if detection occurred on the lower output port as in case (b).}
\end{figure}

A similar reasoning can be made regarding the change in the position of one photon of the output state, from an even mode to the following odd mode (i.e., from mode $2m$ to mode $2m+1$, with the exception of mode $2m=N$ for which has to be exchanged with mode 1).  This pair of odd and even input modes of the interferometer correspond respectively to the upper and lower input modes of one beam splitter of the second layer, looking at the full interferometer. 
Let $\theta_1$ and $\theta_2$ be the phases that a photon, entering respectively the upper and lower input ports of the beam splitter, has already acquired before entering. Upon detection on the upper output mode (Fig.~\ref{fig:exchangeOutput}a), the total phase is $\theta'_1 = \theta_1$ or $\theta'_2=\theta_2+\pi/2$, respectively, depending on the input port. On the other hand, upon detection on the lower output mode (Fig.~\ref{fig:exchangeOutput}b), the total phase at the output is respectively $\theta'_1 = \theta_1+\pi/2$ or $\theta'_2=\theta_2$. If we add suitable phase shifters on the upper input and output modes (Fig.~\ref{fig:exchangeOutput}c), we can make a photon detected on the upper mode acquire the same phases as if it were detected on the lower mode. The relevant phase shifter, with regard to the output photon distribution, is only the one that would be added on the internal arms of the full interferometer, i.e. the one at the input of the considered beam splitter, having a value of $\pi$. Also in this case, the addition of this phase term to the internal phase is equivalent to a change of the sign of $\cos \alpha_1$ in the Eq.~\eqref{eq:Pind}. 

We note that both the exchange operations discussed here above,  when they are performed on a state compatible with Proposition~\ref{prop:allStates},  produce again a state compatible with that Proposition. More precisely, all the input and output states compatible with Proposition \ref{prop:allStates} can be transformed one in the other by a certain number of such exchange operation. 

In particular, all the input states compatible with the Proposition can be derived from the state $\vec g = (1,3,5,\ldots,2n-1)$ with $p$ exchanges of photon position, and analogously all the relevant output states can be obtained from $\vec h = (1,3,5,\ldots,2n-1)$ with $q$ exchanges. Each of these exchanges is associated to a signal change the term $\cos \alpha_1$, resulting in a multiplicative factor $(-1)^{p+q}$. 

One notes that $p$ and $q$ correspond to the number of occupied even modes in the input and output states respectively, and proof of the Proposition is completed.

\section{Experimental details\\ of the photon source}
\label{app:exp}

\begin{figure*}
\centering
\includegraphics[width=0.5\textwidth]{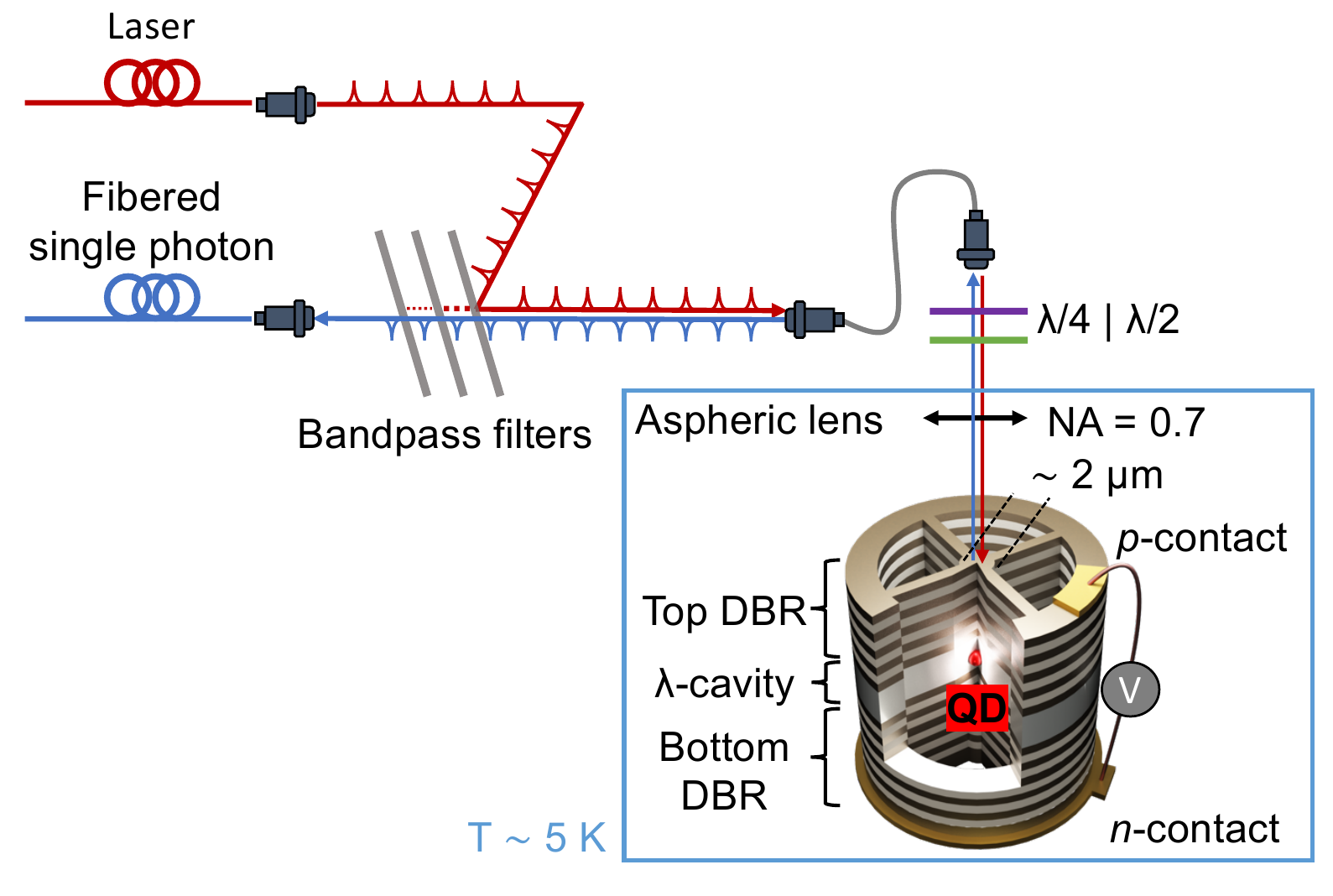}
\caption{Schematic of the experimental setup to generate a single photons stream. (a) The single photon stream is generated by a quantum dot single-photon source based on a neutrally charged InGaAs quantum dot embedded in an electrically-contacted micropillar ($\varnothing$~=~2~$\mu$m) cavity operating at $\sim$5~K. The QDSPS is excited using acoustic-phonon assisted near-resonant ($\Delta \lambda$~=~0.7~nm) excitation \cite{Thomas2021} with a 10 ps pulsed (82~MHz) laser at  $\lambda$=924.4~nm. We apply a voltage of V = -2.0 V to tune the QD into resonance with the cavity. The photons emitted at $\lambda$=925.0~nm are collected with a high numerical aperture (NA = 0.7) aspheric lens mounted inside the cryostat to mitigate mechanical noise. This excitation scheme allows to reach a high polarised first lens brightness ($\sim$19$\%$), defined as the probability per pulse that a single photon is generated and exits the micropillar and reaches the aspheric lens. This is calculated by recording the generated rate of photons and accounting for the optical losses of the setup.
We separate the single photon from the laser using three bandpass filters, resulting in a total suppression of the excitation laser of around 120~dB. This optical setup allows for a high fibered brightness ($\sim9.5\%$).}
\label{fig:sample}
\end{figure*}

The single-photon source is based on a quantum dot coupled to a micropillar cavity. The micropillar is fabricated from a planar sample embedding a $\lambda$-cavity, surrounded by two Distributed-Bragg-Reflectors (GaAs/Al$_{0.9}$Ga$_{0.1}$As, with 14 (28) pairs for the top (bottom) mirror). The $\lambda$-cavity is made of a GaAs, with a single InGaAs QD, and a 20-nm thick tunneling barrier of Al$_{0.1}$Ga$_{0.9}$As positioned 10~nm above the QD layer. The sample is doped to get an effective p-i-n diode structure and the micro-pillar is contacted to a larger structure for electrical contact, as shown in Fig.~\ref{fig:sample}.

The electrically contacted micropillar fabricated using the in-situ lithography technique \cite{dousse2009, Nowak2014a} has a quality factor of Q $\sim$ 2770. Using the Stark effect induced by the electrical field applied to the p-i-n diode we are able to tune the QD in resonance with the micro-cavity, yielding an emission lifetime of T$_1$ = 145~ps and an indistinguishability of $M_\mathrm{s} = (92.3\pm0.1)\%$. The potential difference also induces bending of the energy bands that flush all trapped charges, thus reducing the charge noise and spectral wandering of the emission~\cite{Somaschi2016}.

We show in Table~\ref{tab:VvsT} the measured visibility (indistinguishability) of two photons emitted by the QDSPS as a function of the delay between their emission. The repetition rate of the laser is 82~MHz which means the delay between a photon and the $p^{th}$ subsequent photon is a multiple of 12.2~ns.

Two different time-to-spatial demultiplexer (DMX) are used to generate a 4-photon state at the input of the C.I, i.e. 4 photons synchronized in 4 different fibers, from a single-channel stream of photon at 82~MHz. Practically speaking in both versions the stream of single-photon is sent to a free-space acousto-optic modulator (AOM) driven by a varying radio-frequency (RF) signal acting as a switch between 4 different fibered outputs (see Fig. 2). What differs in the 2 versions is the switching time between two RF frequencies, and the total insertion loss. The first version (DMX4) is a prototype provided by Quandela as a proof-of-concept. The switching time between two RF frequencies, defined as the time to switch from 5\% (channel OFF) to 95\% (channel ON) of maximal transmission, is of the order of 120~ns. The RF driver is set to a predefined frequency corresponding to diffraction to a given output for a duration $\tau$~=~200~ns. The operating time of one channel is thus $\Delta$T~=~320~ns, yielding a maximum delay between two photon of 960~ns. For this first version the total insertion loss (FC/PC fiber to FC/PC fiber) is $\sim$3.5~dB. The second version (DMX6) is a commercially available rackable 6-outputs spatial demultiplexer loaned by Quandela. The design of the DMX has been revisited and optimised to reach a high mechanical stability, a low insertion loss and a fast switching time. The switching time between two RF frequencies has been lowered to 60~ns. We address each output for $\tau$~=~115~ns so that the operating time is $\Delta$T~=~175~ns and the maximal delay between photon is 525~ns. The total insertion loss is $\sim$1.9~dB.

\begin{table*}
    \centering
    \begin{tabular}{||c || c| c |c |c| c||} 
         \hline
         Delay [ns] & 12.2~$\pm$~0.1 & 175~$\pm$~1  & 320~$\pm$~1  & 525~$\pm$~1  & 960~$\pm$~1  \\ [0.5ex]
         \hline
         Emitted photons & 1 & 14 & 26 & 43 & 79 \\ [0.5ex] 
         \hline
		 $g^2$(0) [\%] & $1.9\pm 0.1$ & $1.7\pm 0.1$ & $1.9\pm 0.1$ & $1.7\pm 0.1$ & $1.2\pm 0.1$ \\ [0.5ex] 
         \hline
         $V_\mathrm{HOM}$ [\%] & 88.6~$\pm$~0.1  & 86.4~$\pm$~0.4  & 84.8~$\pm$~0.1  & 76.9~$\pm$~0.4   & 74.2~$\pm$~0.1   \\ 
         \hline
         $M_\mathrm{s}$ [\%] & 92.3~$\pm$~0.2  & 89.6~$\pm$~0.2  & 
         88.4~$\pm$~0.2 & 
         80.0~$\pm$~0.2 & 
         76.0~$\pm$~0.2 \\ 
         \hline
    \end{tabular}
    \caption{\label{tab:VvsT} Visibility (indistinguishability) of two photons emitted by the QDSPS as a function of the delay between their emission. The $g^2(0)$ value is also specified.}
\end{table*}

\begin{figure*}
\centering
\includegraphics[width=0.6\linewidth]{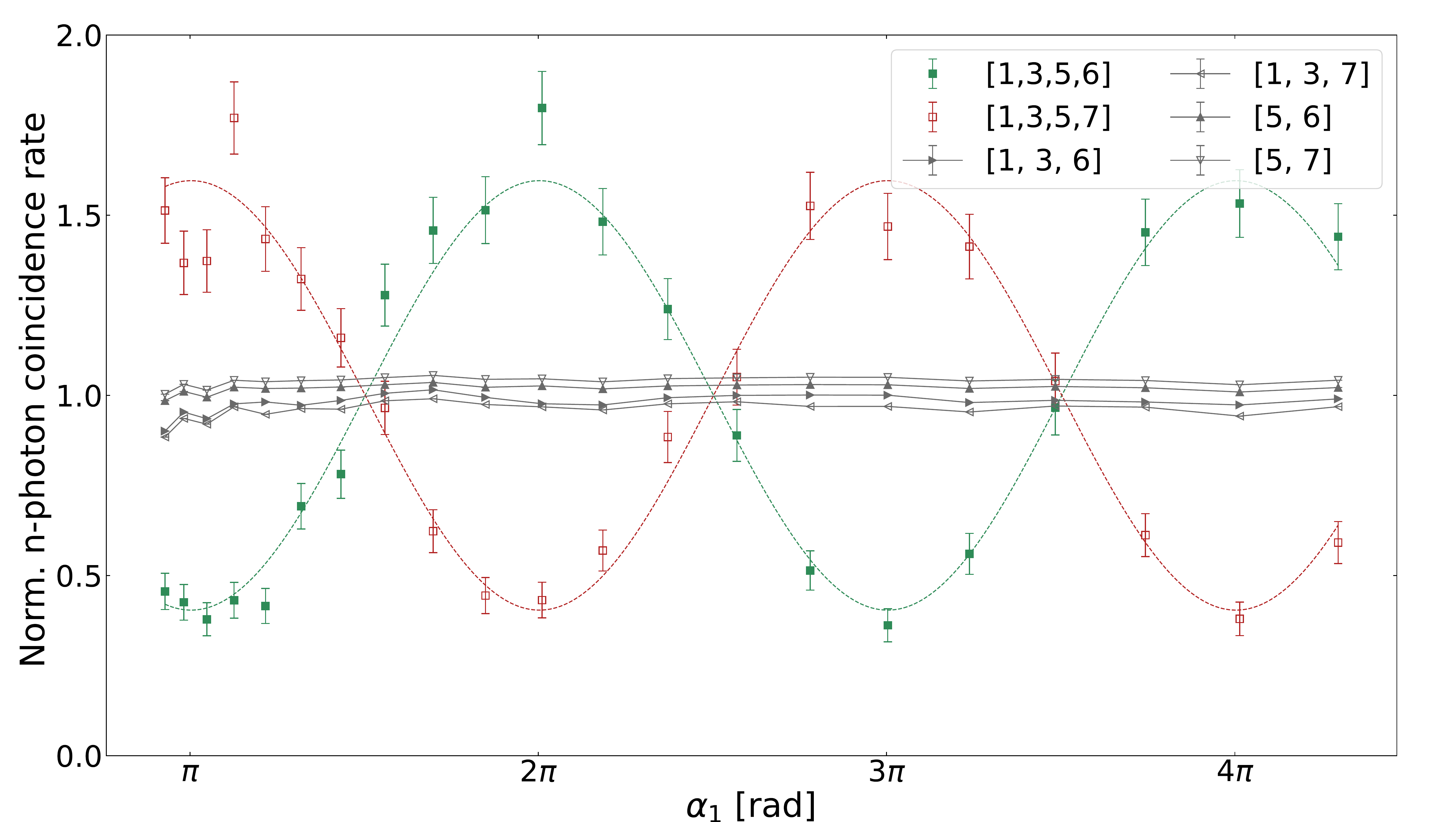}
\caption{\label{fig:Figure_Coincidences} Normalized rate of four-photon coincidences as a function of the internal phase of the C.I., $\alpha_1$, for two output modes corresponding to constructive (destructive) interference, compared to a selection of normalized two- and three-photon coincidences. Note that the two- and three-photon coincidences have been normalised by the mean number of two-, three-photon coincidences respectively, and offset by a constant value so they do not overlap on the plot. The 4-photon interference data in this graph is taken from Fig.~\ref{fig:histo_indist}.}
\end{figure*}

\section{$n<4$-photon coincidences}
\label{app:2&3-coincidences}

We show experimentally the global internal phase of the interferometer only gives rise to interference fringes for four photon coincidences, as expected from the theory (see Proposition \ref{prop:lessPhotons}, Section \ref{sec:interference}).

While acquiring the 4-photon coincidences we also monitor all two- and three-photon coincidences. We select here two particular four-channel output states of interest from Table~\ref{tab:states}, namely (1,3,5,7) and (1,3,5,6), on which we observe clear variation in the four-photon coincidences corresponding to constructive and destructive four-photon interference respectively. We display the rate of two- and three-photon coincidences on various subsets of these channels, as we vary the internal phase: In Fig.~\ref{fig:Figure_Coincidences} we show the normalised three-photon coincidence rate on channels (1,3,6) and (1,3,7), as well as the two-photon coincidence rate on channels (5,6) and (5,7). We observe no variation in the rate of two- and three-photon coincidences with the phase $\alpha_1$, as expected. Note that the same result is observed for all permutations of channels. We only plot a small selection here for clarity.

\section{Bounds for $c_{1}$ and\\ for unmeasured overlaps}
\label{app:bounds}

As discussed in Refs.~\cite{brod2019w,giordani2020,brod2020}, specific families of interferometers can be employed to perform $n$-photon indistinguishability tests based on the measurement of pairwise overlaps $M_{ij}$ between the photons. The overlaps are related to the bunching probabilities $p_{ij}^{\mathrm{b}}$ in a 50/50 beam-splitters as $M_{ij} = 2 p_{ij}^{\mathrm{b}} - 1$, and can be estimated from the Hong-Ou-Mandel visibilities $V_{ij}$. The possibility of bounding the value of $n$-photon indistinguishability is obtained by considering that certain families of interferometers can be described as graphs. More specifically, in interferometers performing measurement of pairwise HOM interference, nodes on the graphs can be associated to the different photons, while edges between two nodes correspond to the measurement of a HOM visibility between the corresponding particles. By considering a state model such as the one of Eq.~\eqref{eq:decomp}, logical propositions associated to the graph structure can be used to provide bounds on $c_1$ based on the measured overlaps. Notably, this methodology can be applied to the interferometer implemented in this paper which, besides allowing for four-photon interference fringes, also performs two-photon HOM interference between certain pairs of input photons. By using the approach of Refs.~\cite{brod2019w,giordani2020,brod2020}, the graph edges can be mapped to logical propositions, which must satisfy appropriate set of inequalities to represent a consistent set of probabilities. We obtain the following bounds for $c_1$ based on the four overlaps for particles A, B, C, D according to the notation in Fig.~3 $(M_{\mathrm{AB}}, M_{\mathrm{BC}}, M_{\mathrm{CD}}, M_{\mathrm{DA}})$ measured via the C.I.:
\begin{align}
c_1 &\geq M_{\mathrm{AB}} + M_{\mathrm{BC}} + M_{\mathrm{CD}} + M_{\mathrm{DA}} - 3 \\
c_1 &\leq \min(M_{\mathrm{AB}}, M_{\mathrm{BC}}, M_{\mathrm{CD}}, M_{\mathrm{DA}})
\end{align}
\begin{align}
M_{\mathrm{AB}} + M_{\mathrm{BC}} - 1 \leq &M_{\mathrm{AC}} \leq 1 - \vert M_{\mathrm{AB}} - M_{\mathrm{BC}} \vert \\
M_{\mathrm{BC}} + M_{\mathrm{CD}} - 1 \leq &M_{\mathrm{BD}} \leq 1 - \vert M_{\mathrm{BC}} - M_{\mathrm{CD}} \vert 
\end{align}

To estimate the upper and lower bounds from experimentally estimated overlaps, that are affected by measurements error, we have applied a bootstrapping approach. More specifically, we generate $10^4$ sets of the four indistinguishability values with a normal distribution matching the experimentally obtained mean and standard deviation, and then calculate the upper and lower bound for each set $\{{c_1}_\mathrm{min}(i), {c_1}_\mathrm{max}(i)\}$ using Eqs~\eqref{eq:c1_lower_bound} and \eqref{eq:c1_upper_bound} respectively. We then take the lower bound as  ${c_1}_\mathrm{min} = \mathrm{mean}\{{c_1}_\mathrm{min}(i)\} - 3 \, \mathrm{std}\{{c_1}_\mathrm{min}(i)\}$, ${c_1}_\mathrm{max} = \mathrm{mean}\{{c_1}_\mathrm{max}(i)\} + 3 \, \mathrm{std}\{{c_1}_\mathrm{max}(i)\}$.

\section{Modeling experimental imperfections}
\label{app:model}

\begin{figure*}
\centering
\includegraphics[width=0.99\textwidth]{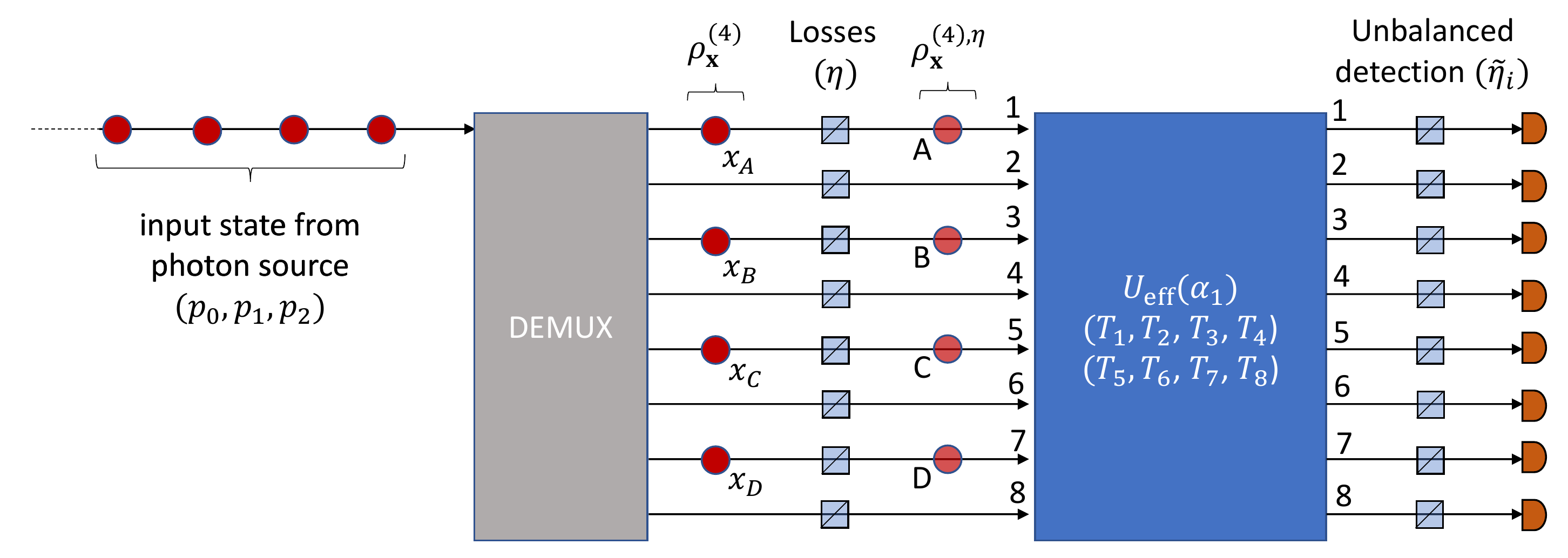}
\caption{Schematic and synoptic view on the parameters involved to model the action of multiphoton emission, circuit errors, losses  and unbalanced detection on the performed experiment to retrieve indistinguishability $c_1$. The modeling of each noise contribution and their respective relevant parameters  are detailed in the text of Appendix~\ref{app:model}. The estimated values $c_1^{\text{mod}}$ turning on successive noise contributions in the model are given in Table~\ref{tab:2} which also specifies the values of the parameters used in each case.}
\label{fig:1}
\end{figure*}

Here we discuss a model to take into account the effect of experimental imperfections in the estimation of $c_1$. In particular, we consider different effects that affect the measured value of $c_1$ following the method described in the main text.
A schematic representation of the full model is shown in Fig.~\ref{fig:1} that displays the relevant parameters that we detail and explain below. The main source of deviation of the estimate from the true $c_1$ value is multiphoton emission, i.e. the non-zero probability that the source emits more than one photon per mode ($p_2$). We then consider and include the effect of losses ($\eta$) within the apparatus, the presence of fabrication errors in the integrated device ($U_{\mathrm{eff}}(\alpha_1))$, and unbalanced detection efficiencies ($\eta_i$). Note that within the model we have also included the effect of partial photon distinguishability ($\rho^{(4)}_x$), which is the actual physical effect that the experiment aims at reconstructing via the estimation of parameter $c_1$. Detailed description on how these effects are included in the model is reported below.

\begin{table}
\begin{tabular}{|c|c|}
\hline
Input configuration & Output configuration \\
\hline
(1,0,1,0,1,0,1,0) & (1,0,1,0,1,0,1,0) \\
\hline
(2,0,1,0,1,0,0,0) & \multirow{3}{*}{(1,0,1,0,1,0,1,0)} \\
+ permutations &\\
on modes (1,3,5,7) &\\
\hline
(2,0,2,0,0,0,0,0) & \multirow{3}{*}{(1,0,1,0,1,0,1,0)} \\
+ permutations &\\
on modes (1,3,5,7) &\\
\hline
(2,0,1,0,1,0,1,0) & (2,0,1,0,1,0,1,0) \\
+ permutations & + permutations\\
on modes (1,3,5,7) & on modes (1,3,5,7)\\
\hline
(2,0,2,0,1,0,0,0) & (2,0,1,0,1,0,1,0) \\
+ permutations & + permutations\\
on modes (1,3,5,7) & on modes (1,3,5,7)\\
\hline
\end{tabular}
\caption{Configurations providing contributions to the input-output probability, in the presence of multiphoton emission and losses, for output modes (1,3,5,7). Analogous tables can be constructed for each set of 4 output modes.}
\label{tab:1}
\end{table}

\begin{table*}
\begin{tabular}{c|c|c|c|c}
$c_{1}^{\mathrm{mod}}$ & Partial distinguishability & Multiphoton terms & Imperfect DC & Unbalanced detection\\
\hline
$0.661 \pm 0.006$ & Yes & No & No & No\\
$0.592 \pm 0.005$ & Yes & Yes & No & No\\
$0.591 \pm 0.005$ & Yes & Yes & Yes & No\\
$0.590 \pm 0.005$ & Yes & Yes & Yes & Yes
\end{tabular}
\caption{\label{tab:2} Results of the numerical simulation for the experiment modeling. In particular, we have progressively added noise contributions for the configuration detailed in Section \ref{sec:exp_model} leading to an experimental value of $c_1 = 0.61 \pm 0.01$. The corresponding parameters for such simulation are reported here. Brightness: $B \sim 0.098$. Second-order correlation: $g^{(2)}(0) = 0.019 \pm 0.001$. Directional coupler DC transmissivities: $T_1 \sim 0.503$, $T_2 \sim 0.508$, $T_3 \sim 0.505$, $T_4 \sim 0.507$, $T_5 \sim 0.506$, $T_6 \sim 0.512$, $T_7 \sim 0.5045$, $T_8 \sim 0.534$. Indistinguishability parameters: $x_{\mathrm{A}} \sim 0.852$, $x_{\mathrm{B}} \sim 0.883$, $x_{\mathrm{C}} \sim 0.941$, $x_{\mathrm{D}} \sim 0.932$. Detection imbalance: $\tilde{\eta_{1}} = 0.92$, $\tilde{\eta_{2}} = 0.90$, $\tilde{\eta_{3}} = 0.92$, $\tilde{\eta_{4}} = 0.91$, $\tilde{\eta_{5}} = 0.90$, $\tilde{\eta_{6}} = 0.90$, $\tilde{\eta_{7}} = 0.90$, $\tilde{\eta_{8}} = 0.90$. Effective losses: $\eta \sim 0.25$ are included in all simulations. Similar analyses are shown in Section~\ref{sec:exp_dist}, Fig.~\ref{fig:Simulations_Bounds}, for other indistinguishability configurations.}
\end{table*}

\textit{Multiphoton emission and losses. --} Let us consider that the photon source has a non-zero $g^{(2)}(0)$. The state generated by the source, on each time-bin (separated by 12.2~ns) before demultiplexing, can be written as:
\begin{equation}
\label{eq:state_1}
\rho_i = p_0 \vert 0_{i} \rangle \langle 0_{i} \vert + p_1 \vert 1_{i} \rangle \langle 1_{i} \vert + p_2 \vert (1,\tilde{1})_{i} \rangle \langle (1,\tilde{1})_{i} \vert.
\end{equation}
Here, $\vert k_{i} \rangle $ stands for $k$ photons on mode $i$, while $p_0$,  $p_1$ and  $p_2$ are the probabilities of having 0, 1, 2 photons in a single time-bin respectively. Their values can be obtained from the source brightness $B \sim p_1 + p_2$ and from the $g^{(2)}(0)$ parameter as $g^{(2)}(0) = 2 p_2/(p_1 + 2 p_2)^2$ by direct calculation from the definition $g^{(2)}(0) = \langle n (n-1) \rangle/\langle n \rangle^{2}$. Notation $\vert (1,\tilde{1})_{i} \rangle$ describes the emission of two photons in the same time-bin, corresponding to the addition of a noise photon to the principal one. Within this model, we will neglect higher order noise terms. For quantum dot sources~\cite{Ollivier2021}, the overlap of such noise photon with the principal one can be approximated to $\sim 0$, that is, the noise photon $(\tilde{1})$ is distinguishable from the others ($1$)~\cite{Ollivier2021}. The complete input state after demultiplexing can be written as a density matrix:
\begin{equation}
\rho^{(4)} = \rho_1 \rho_3 \rho_5 \rho_7,
\end{equation}
where $\rho_i$ is the incoherent mixture of Eq.~\eqref{eq:state_1}. For the low values of $g^{(2)}(0)$ attained by the source described in this paper, $p_2$ is small with respect to $p_1$. It is then possible to neglect all terms with more than one noise photon in state $\rho^{(4)}$. By further keeping only terms with at least 4 photons (that are the only relevant ones for the four-photon coincidence measurements), the density matrix can thus approximated as:
\begin{equation}
\begin{aligned}
\rho^{(4)} &\sim p_{1}^{4} \vert 1_{1}, 1_ {3}, 1_{5}, 1_{7} \rangle \langle 1_{1}, 1_ {3}, 1_{5}, 1_{7} \vert +\\
&+ p_{0} p_1^{2} p_{2} \Big\{ \vert (1,\tilde{1})_{1}, 0_ {3}, 1_{5}, 1_{7} \rangle \langle (1,\tilde{1})_{1}, 0_ {3}, 1_{5}, 1_{7} \vert +\\
&+ \vert (1,\tilde{1})_{1}, 1_ {3}, 0_{5}, 1_{7} \rangle \langle (1,\tilde{1})_{1}, 1_ {3}, 0_{5}, 1_{7} \vert + \ldots \Big\} +\\
&+ p_{1}^{3} p_{2} \Big\{ \vert (1,\tilde{1})_{1}, 1_{3},1_{5},1_{7} \rangle \langle (1,\tilde{1})_{1},1_{3},1_{5},1_{7} \vert +\\
&+ \vert 1_{1},(1,\tilde{1})_{3},1_{5},1_{7} \rangle \langle 1_{1},(1,\tilde{1})_{3},1_{5},1_{7} \vert + \ldots \Big\},
\end{aligned}
\end{equation}
where all possible permutations have to be included in the parentheses.

Up to now we have neglected the effect of losses. If losses are almost equally distributed between each arm of the interferometer, we can apply the results of Ref.~\cite{Oszm18}. In this case, losses commute with linear optical elements, including demultiplexing and detection efficiencies. Thus, we can equivalently put all the losses occurring in the apparatus right before the input of the 8-mode interferometer, by defining the overall transmission parameter $\eta$ (per photon). The state after losses can be written as the sum of different contributions:
\begin{widetext}
\begin{equation}
\label{eq:4-mode}
\begin{aligned}
\rho^{(4),\eta} &= \big[p_{1}^{4} \eta^{4} + 4 p_{1}^{3} p_{2} \eta^{4} (1-\eta) \big] \vert 1_{1}, 1_ {3}, 1_{5}, 1_{7} \rangle \langle 1_{1}, 1_ {3}, 1_{5}, 1_{7} \vert+ \\ 
&+\big[ p_{1}^{3} p_{2} \eta^{4} (1 - \eta) \big] \left\{ \vert \tilde{1}_{1}, 1_ {3}, 1_{5}, 1_{7} \rangle \langle \tilde{1}_{1}, 1_ {3}, 1_{5}, 1_{7} \vert + \vert 1_{1}, \tilde{1}_ {3}, 1_{5}, 1_{7} \rangle \langle 1_{1}, \tilde{1}_ {3}, 1_{5}, 1_{7} \vert + \ldots \right\} +\\
&+ \big[p_{0} p_{1}^{2} p_{2} \eta^{4} + p_{1}^{3} p_{2} \eta^{4} (1-\eta) \big] \left\{ \vert (1,\tilde{1})_{1}, 0_{3}, 1_{5}, 1_{7} \rangle \langle (1,\tilde{1})_{1}, 0_{3}, 1_{5}, 1_{7} \vert + \vert (1,\tilde{1})_{1}, 1_{3}, 0_{5}, 1_{7} \rangle \langle (1,\tilde{1})_{1}, 1_{3}, 0_{5}, 1_{7} \vert + \ldots \right\} + \\
&+ \big[ p_{1}^{3} p_{2} \eta^{5} \big] \left\{\vert (1,\tilde{1})_{1}, 1_{3}, 1_{5}, 1_{7} \rangle \langle (1,\tilde{1})_{1}, 1_{3}, 1_{5}, 1_{7} \vert + \vert 1_{1}, (1,\tilde{1})_{3}, 1_{5}, 1_{7} \rangle \langle 1_{1}, (1,\tilde{1})_{3}, 1_{5}, 1_{7} \vert + \ldots \right\}
\end{aligned}
\end{equation}
\end{widetext}
where the states are written according to the notation described above.

The first three groups of terms correspond to states with 4 input photons on the device, and will provide a contribution to the detection of a given 4-photon event whenever the corresponding probability is non-zero. Conversely, the last terms correspond to 5 input photons on the device. By considering that SNSPDs are non-photon number resolving detectors, it is necessary to consider that they can provide non-zero contribution to 4-detector clicks in different cases. For instance, let us take output modes (1,3,5,7). A non-zero contribution is obtained for the 5-photon terms when one of the following output configuration is obtained:
\begin{equation}
\label{eq:permut_5photon}
\begin{aligned}
&(2,0,1,0,1,0,1,0); (1,0,2,0,1,0,1,0); \\
&(1,0,1,0,2,0,1,0); (1,0,1,0,1,0,2,0).
\end{aligned}
\end{equation}
where $(i_1, i_2, \ldots, i_8)$ stand for $i_k$ photons on output mode $k$ (see the second line of Table~\ref{tab:1}). 

The overall output probability is finally retrieved by summing up all probabilities corresponding to terms and weights of Eq.~\eqref{eq:4-mode}, by considering for 5-photon terms all possibile configurations yielding to a useful signal according to Eq. \eqref{eq:permut_5photon}. 

\textit{Partial photon distinguishability. --} To include the effect of partial photon distinguishability, one can rely on different papers discussing such effect in a Boson Sampling framework \cite{tichy2015,Rene18,Moyl19}. These models substantially take into account that indistinguishability between the particles is described by an Hermitian matrix $\mathcal{S}_{ij}$, representing the set of pairwise overlaps. In general, one has $\vert \mathcal{S}_{ij} \vert \leq 1$ (for $i \neq j$), while $\mathcal{S}_{ij} =1$ is obtained when photons $i$ and $j$ are indistinguishable. Such overlaps can be different for each photon pair, and include all degrees of freedom that add distinguishability to the generated photon state. Notably, HOM pairwise visibility between particles $i$ and $j$ provides information on the moduli $\vert \mathcal{S}_{ij} \vert$, while it is insensitive to the complex phases.

We have assumed, in our model, the simplified scenario where $\mathcal{S}_{ij}$ are real numbers. Furthermore, we have considered a description where each principal photon has a probability $x_i$ to be indistinguishable, and a probability $(1 - x_i)$ to be fully distinguishable from the others. As shown in the Main Text, this specific choice is able to provide an accurate description of our multi-photon experiment, performed with the QDSPS and the C.I. device. Thus, the density matrix of the input state $\rho^{(4)}$, and $\rho^{(4),\eta}$ after including the effective action of losses, has to be replaced with effective states $\rho^{(4)}_{\mathbf{x}}$ and  $\rho^{(4),\eta}_{\mathbf{x}}$, where $\mathbf{x}$ is the vector of parameters $(x_\mathrm{A}, x_\mathrm{B}, x_\mathrm{C}, x_\mathrm{D})$. Conversely, noise photons are considered fully distinguishable, thus corresponding to a value $\tilde{x} = 0$.

\textit{Circuit parameters. --} Circuit errors can be introduced in the presence of fabrication imperfections. The relevant error to be considered in the model corresponds to directional couplers with transmissivities different from the expected value $T_j = 0.5$. Losses can be included in the parameter $\eta$ discussed above. The values of the different transmissivities $T_j$ for the 8 directional couplers implemented in the structure have been characterized before the experiment, and are reported in Tab.~\ref{tab:2}. Their values can be used to correct the effective unitary transformation  $U_{\mathrm{eff}}(\alpha_1)$ implemented by the integrated device. This effective matrix is used to calculate the output probabilities in the model.

\textit{Unbalanced detection efficiencies. --} Finally, we can include the action of unbalanced detection efficiencies. This is performed by considering that detection efficiency for each output mode can be written as $\eta_i = \eta_0 \tilde{\eta}_i$, where $\eta_0$ is equal for all the modes, while $\tilde{\eta}_i$ represents the imbalance. Note that by construction $\mathrm{max}_{i} \tilde{\eta}_i = 1$. Starting from this equation, we observe that $\eta_0$ is a common set of balanced losses that can be included within the parameter $\eta$ discussed above. Conversely, the imbalance $\tilde{\eta}_{i}$ are included by considering that each detector clicks with probability $p_{i}(n_{i}) = 1 - (1-\tilde{\eta}_{i})^{n_i}$, where $n_{i}$ is the number of impinging photons. Output probabilities obtained from the calculations can be thus corrected accordingly, by adding the imbalance in the transition probabilities from the input states of the density matrix expansion in Eq.~\eqref{eq:4-mode} and the corresponding output configuration.

\textit{Numerical simulations. --} Starting from the model described above, we have performed some numerical simulations to investigate how such errors affect the measurement of $c_1$. Such a parameter $c_1$ is indeed retrieved in the experiment from the visibility of the output probabilites as a function of $\alpha_1$, that is, the internal phase of the interferometer. In particular, in the experiment we have summed up all $8$ output configurations varying as $1 + c_1 \cos(\alpha_1)$, and all $8$ terms changing as varying as $1 - c_1 \cos(\alpha_1)$ to obtain the final estimate of $c_1$. We have thus performed a full numerical simulation of the experiment, which allowed us to evaluate the contribution of each term to the measured value of $c_1$.

As a first step, parameters $p_0, p_1, p_2$ have been obtained by considering the actual source brightness $B$, and the value of $g^{(2)}(0)$. Regarding losses, transmission parameter $\eta$ has been chosen to obtain the measured single-photon count rate (thus including an effective transmission parameter for the demultiplexing stage). Conversely, the imbalance $\tilde{\eta}_{i}$ in detection efficiencies and the directional coupler transmittivities have been directly calibrated in the setup.

The first stage of the simulation corresponds to finding an estimate of parameters $x_{i}$ describing partial photon distinguishabilities for modes $(\mathrm{A}, \mathrm{B}, \mathrm{C}, \mathrm{D})$. This can be performed by considering that those parameters are strictly related to the two-photon overlaps $M_{ij} = x_i x_j$, and can thus be retrieved from the measured visibilities $V_{ij} = (V_{\mathrm{AB}}, V_{\mathrm{BC}}, V_{\mathrm{CD}}, V_{\mathrm{DA}})$, by including the effect of $g^{(2)}(0)$ to find the instrinic indistinguishability $M_{ij}$ as discussed in Ref.~\cite{Ollivier2021}, and by correcting for imperfect directional couplers.

Then, we have performed a full simulation of the experiment for the estimate of $c_1$ according to the model above. As an example, in Table~\ref{tab:2} we report the effect of each contribution on one of the experimental points reported in the paper, corresponding to $V_\mathrm{AB} = 0.727 \pm 0.001$ ($M_\mathrm{AB} = 0.760 \pm 0.002$), $V_\mathrm{BC} = 0.790 \pm 0.001$, ($M_\mathrm{BC} = 0.825 \pm 0.002$), $V_\mathrm{CD} = 0.848 \pm 0.001$, ($M_\mathrm{CD} = 0.884 \pm 0.002$), $V_\mathrm{DA} = 0.755 \pm 0.002 $, ($M_\mathrm{DA} = 0.789 \pm 0.003 $), corresponding to an extracted value of $c_1 = 0.61 \pm 0.01$. By including all effects, the predicted value of $c^{\mathrm{mod}}_1$ is $c^{\mathrm{mod}}_1 \sim 0.590 \pm 0.005$, in good agreement with the experimental measurement. For this value, the major contribution $\sim 0.07$ derives from multiphoton emission, while a minor correction is provided by imperfect directional couplers (DCs) ($\sim 0.001$) and unbalanced detection efficiencies ($\sim 0.001$).

\end{document}